\documentclass[12pt, a4paper]{article}
\usepackage[latin1]{inputenc}
\usepackage{vmargin}
\usepackage{amsfonts}
\usepackage{amssymb}
\usepackage{epstopdf}
\usepackage[dvips]{graphicx}
\usepackage{graphicx}
\usepackage{amsmath}
\usepackage{amsthm}
\usepackage{setspace}
\usepackage{multirow}
\usepackage{hyperref}
\usepackage{cases}
\usepackage[bottom]{footmisc}
\usepackage[small]{caption}
\usepackage{lscape}

\setmarginsrb{1.8cm}{1.8cm}{1.8cm}{1.8cm}{0cm}{0cm}{0cm}{0.5cm}

\begin{document}

\newtheorem{Proposition}{Proposition}
\newtheorem{Theorem}{Theorem}
\newtheorem{Corollary}{Corollary}
\newtheorem{Def}{Definition}
\newtheorem{Lem}{Lemma}

\title{Price jump prediction in Limit Order Book}
\author{Ban Zheng\addtocounter{footnote}{0}\thanks{Natixis, Equity Markets. \texttt{ban.zheng@natixis.com}. The authors would like to thank the members of Natixis quantitative research team for fruitful discussions}, Eric Moulines\addtocounter{footnote}{5}\thanks{Departement TSI, T\'el\'ecom ParisTech. \texttt{eric.moulines@telecom-paristech.fr}}, Fr\'ed\'eric Abergel\addtocounter{footnote}{4}\thanks{BNP Paribas Chair of Quantitative Finance, Ecole Centrale Paris, MAS Laboratory. \texttt{frederic.abergel@ecp.fr}}}
\date{March 2012}

\maketitle
\abstract{A limit order book provides information on available limit order prices and their volumes. Based on these quantities, we give an empirical result on the relationship between the bid-ask liquidity balance and trade sign and we show that liquidity balance on best bid/best ask is quite informative for predicting the future market order's direction. Moreover, we define price jump as a sell (buy) market order arrival which is executed at a price which is smaller (larger) than the best bid (best ask) price at the moment just after the precedent market order arrival. Features are then extracted related to limit order volumes, limit order price gaps, market order information and limit order event information. Logistic regression is applied to predict the price jump from the limit order book's feature. LASSO logistic regression is introduced to help us make variable selection from which we are capable to highlight the importance of different features in predicting the future price jump. In order to get rid of the intraday data seasonality, the analysis is based on two separated datasets: morning dataset and afternoon dataset. Based on an analysis on forty largest French stocks of CAC40, we find that trade sign and market order size as well as the liquidity on the best bid (best ask) are consistently informative for predicting the incoming price jump.
}

\section*{Introduction}
The determination of jumps in financial time series already has a long history as a challenging, theoretically interesting and practically important problem. Be it from the point of view of the statistician trying to separate, in spot prices, those moves corresponding to "jumps" from those who are compatible with the hypothesis of a process with continuous paths, or from the point of view of the practitioner: market maker, algorithmic trader, arbitrageur, who is in dire need of knowing the direction and the amplitude of the next price change, there is a vast, still unsatisfied interest for this question.
Several attempts have been made at theorizing the observability of the difference between processes with continuous or discontinuous paths, and the major breakthrough in that direction is probably due to Barndorff-Nielsen and Shephard {~\cite{BNS2004}}, who introduced the concept of bi-power variation, and showed that - in a nutshell - the occurrence of jumps can be seen in the limiting behavior as the time step goes to zero of the bi-power variation: for a process with continuous paths, this quantity should converge to (a multiple of) the instantaneous variance, and the existence of a possibly different limit will be caused by the occurrence of jumps.

Since then, many authors, in particular A\"{\i}t-Sahalia and Jacod (2009){~\cite{ASJ2009}} have contributed to shed a better light on this phenomenon, and one can safely say that rigorous statistical tests for identifying continuous-time, real-valued processes with discontinuous paths are now available to the academic community as well as the applied researcher.

However, it is a fact that the physics of modern, electronic, order-driven markets is not easily recast in the setting of real-valued, continuous-time processes, and it is also a fact that the time series of price, no matter how high the sampling frequency, is not anymore the most complete and accurate type of information one can get from the huge set of financial data at our hands. In fact, a relatively recent trend of studies has emerged over the past 10 years, where the limit order book became the center of interest, and the price changes are but a by-product of the more complicated set of changes on limit orders, market orders, cancellation of orders, ... see e.g. Chakraborti et al. (2009) {~\cite{Chakraborti2009}}, Abergel et al. (2011) {~\cite{ACCM2011}} for the latest developments in the econophysics of order-driven markets. This new standpoint is quite enlightening, in that the physics of price formation becomes much more apparent, but it calls for a drastic change in the basic modeling tools: prices now live on a discrete grid with a step size given by the tick, the changes in price occur at discrete times. Furthermore, a host of important events that affect the order book rather than the price itself, events which are therefore essential in understanding the driving forces of the price changes, now become observable, and their role in the price dynamics must be taken into account when one is interested in understanding the latter.

Our point of view is slightly different: rather than concentrate on the one-dimensional price time series, we want to model the dynamics in event time of the whole order book, and focus on some specific events that can be interpreted in terms of jumps. To do so, we shall depart from the classical definitions - if any such thing exists - of a jump in a financial time series, and restrict ourselves to the more natural, more realistic and also more prone to experimental validation, concept of a \emph{inter-trade price jump} and \emph{trade-through}.

By definition, an \emph{inter-trade price jump} is defined as an event where a market order is executed at a price which is smaller (larger) than the best limit price on the Bid (Ask) just after the precedent market order arrival. An \emph{inter-trade price jump} permits a limit order submitted at the best bid (best ask) just after a market order arrival to be surely executed by the next market order arrival. A \emph{trade-through} corresponds to the arrival of a new market order, the size of which is larger than the quantity available at the best limit on the Bid (for a sell order) or Ask (for a buy order) side of the order book. By nature, such an order will imply an automatic and instantaneous price change, the value of which will be exactly the difference in monetary units between the best limit price before and after transaction on the relevant side of the order book. \emph{Trade-through} can be interpreted as the instantaneous price change triggered by a market order, meanwhile, \emph{inter-trade price jump} is post-trade market impact. Most of researches on limit order book are based on stocks and often relates to characterizing features such as liquidity, volatility and bid-ask spread instead of making prediction, see Hasbrouck (1991) {~\cite{Hasbrouck1991}}, Hausman et al. (1992) {~\cite{HLM1992}}, Keim and Madhavan (1996) {~\cite{KM1996}}, Lo et al. (2002) {~\cite{LMCZ2002}}, Lillo et al. (2003) {~\cite{LFM2003}}, Hasbrouck (2006) {~\cite{Hasbrouck2006}}, Parlour and Seppi (2008){~\cite{PS2008}} and Jondeau et al. (2008) and Linnainmaa and Rosu (2009) {~\cite{LR2009}}. \emph{Trade-through} has also been the object of several recent studies in the econometrics and finance literature, see e.g. Foucault and Menkveld (2008) {~\cite{FM2008}}(for cross-sectional relationship study) and Pomponio and Abergel (2011) {~\cite{PA2011}}.

In this work, we investigate whether the order book shape is informative for the \emph{inter-trade price jump} prediction and whether \emph{trade-through} contributes to this prediction. Recently, many researchers propose machine learning methods to make prediction on limit order book. Blazejewski and Coggins (2005){~\cite{Blazejewski2005}} present a non-parametric model for \emph{trade sign} (market order initiator) inference and they show that limit order book shape and historical trades size are informative for the \emph{trade sign} prediction. Fletcher et al. (2010) {~\cite{FLETCHER2010}} applied multi kernel learning with support vector machine in predicting the EURUSD price evolution from the limit order book information. Here, \emph{logistic regression} is introduced to predict the occurrence of \emph{inter-trade price jump}. Variable selection by \emph{lasso logistic regression} provides us an insight into the dynamics of limit order book and allows us to select the most informative features for predicting relevant events. We will show that some features of the limit order book have strong predictive and explanatory power, allowing one to make a sound prediction of the occurrence of \emph{inter-trade price jump} knowing the state of the limit order book. \emph{Trade-through} is also confirmed to be quite informative for \emph{inter-trade price jump} prediction. This result in itself is interesting in that it allows one to use the full set of available information in order to do some prediction: whereas the history of the price itself is known not to be a good predictor of the next price moves - the so-called efficiency of the market is relatively hard to beat when one only uses the price information - we shall show that the limit order volumes contain more information, and the market order size contributes also to an accurate prediction of \emph{inter-trade price jump}.

This paper is organized as follows. Section ~\ref{section: Description} describes the main notations in limit order book. Section~\ref{section: EmpiricalFacts} gives an empirical result on the relationship between Bid-Ask liquidity balance and trade sign. Section ~\ref{section: Logit} introduces \emph{logistic regression} for \emph{inter-trade price jump} prediction and \emph{lasso logistic regression} for variable selection. The conclusion is in Section ~\ref{section: Conclusion}.

\section{Description and data notation}
\label{section: Description}
\begin{figure}[!h]
  \center
\includegraphics[trim=5mm 5mm 5mm 5mm, clip, height=10cm, width=4.2cm, angle=-90]{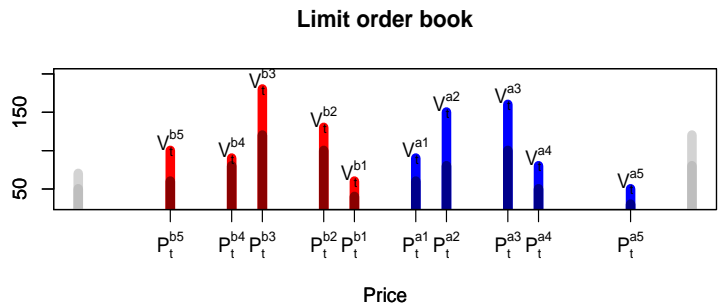}
\caption{Limit Order Book description. Limit order price is discretized by tick price.}
\label{LOB}
\end{figure}

The Euronext market adopts NSC (Nouveau Syst\`eme de Cotation) for electronic trading. During continuous trading from $9h00$ to $17h30$, NSC matches market orders against the best limit order on the opposite side. Various order types are accepted in NSC such as limit orders (an order to be traded at a fixed price with certain amount), market orders (order execution without price constraint), stop orders (issuing limit orders or market orders when a triggered price is reached) and iceberg orders (only a part of the size is visible in the book). Limit order is posted to electronic trading system and they are placed into the book according to their prices, see Figure {~\ref{LOB}}. Market order is an order to be executed at the best available price in limit order book. The lowest price of limit sell orders is called \emph{Best Ask}; the highest price of limit buy orders is called \emph{Best Bid}. The gap between the \emph{Best Bid} and the \emph{Best Ask} is called the \emph{Spread}. When a \emph{market buy order} with price higher/equal than the \emph{best ask price}, a trade occurs and the limit order book is updated accordingly. Limit orders can also be cancelled if there have not been executed, so the limit order book can be modified due to limit order cancellation, limit order arrival or market order arrival. In case of iceberg orders, the disclosed part has the same priority as a regular of limit order while the hidden part has lower priority. The hidden part will become visible as soon as the disclosed part is executed. The case that the hidden part is consumed by a market order without being visible before is quite rarely. In this study, we neglect stop orders and iceberg orders which are relatively rare compared to limit order and market order events.

In a limit order book, as shown in figure $~\ref{LOB}$, only a certain number of best buy/sell limit orders are available for public. We denote the number of available bid/ask limit prices by $L$.

\begin{figure}[!h]
\center
\includegraphics[trim=0mm 0mm 0mm 0mm, clip, height=12cm, width=12cm, angle=-90]{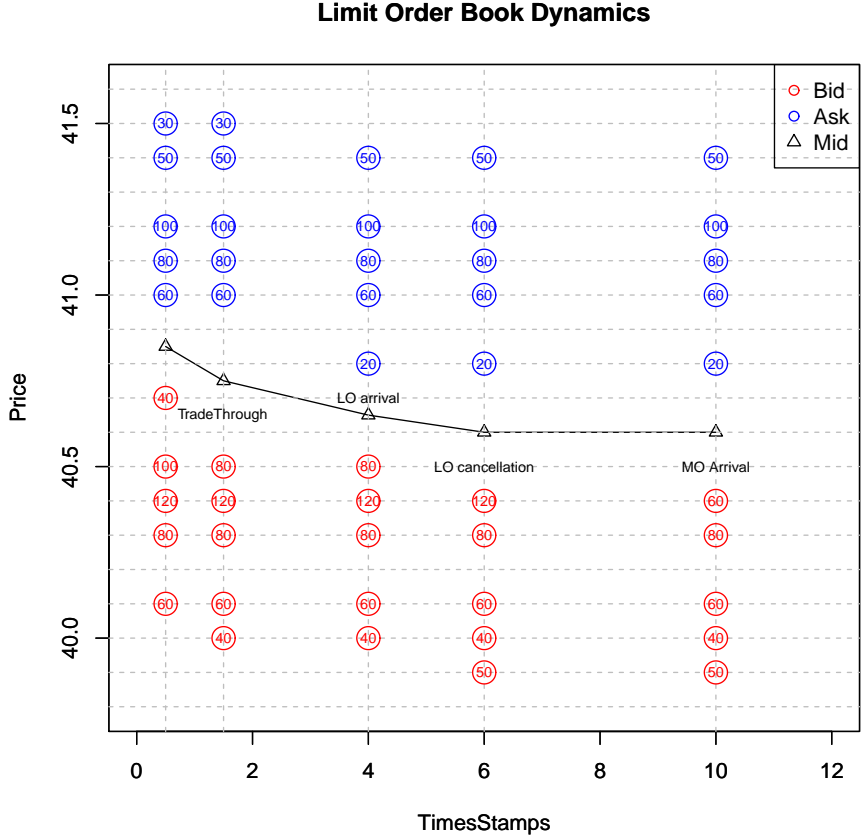}
\caption{Dynamics of limit order book. The first event is a \emph{trade-through} event where a market order consumes $60$ stocks at the bid side, then a new ask limit order of size $20$ arrives in the Spread. Successively, a cancellation at the best bid price follows and the precedent second best bid price emerges the best bid price, the a regular market order triggers a transaction of size $60$.}
\label{LOB_Dynamics}
\end{figure}

In this study, for simplicity, we focus on limit order arrival events, limit order cancellation events and market order arrival events, see Figure {~\ref{LOB_Dynamics}}. The number of visible limit order levels is chosen to be five $L = 5$. Our dataset is provided by NATIXIS via Thomson Reuter's `Reuters Tick Capture Engine' and comprises of trades and limit order activities of the $40$ member stocks of index $CAC40$ between April 1st 2011 and April 30th 2011. In order to get rid of open hour and close hour, we extract the data from $09h05$ to $17h25$. Every transaction and every limit order book modification are recorded in milliseconds. The data contains information on the $L$ best quotes on both bid and ask sides. The trade data and quotes data are matched.

Denote t as a time index indicating all limit order book events. $P_t^{b, i}$ and $P_t^{a, i}$ for $i = 1, \cdots, L$ define the $i^{th}$ best log bid/ask quote instantaneously after the $t^{th}$ event. We denote $S_t = P_t^{a, 1} - P_t^{b, 1}$ the spread instantaneously after the $t^{th}$ event. $G_t^{b, i} = P_t^{b, i} - P_t^{b, i+1}$, $G_t^{a, i} = P_t^{a, i+1} - P_t^{a, i}$ for $i = 1, \cdots, L-1$ define respectively the $i^{th}$ best bid(ask) limit price gap instantaneously after the $t^{th}$ event. Besides, $V_t^{b, i}$ and $V_t^{a, i}$ for $i = 1, \cdots, L$ denote the log limit order volume on the $i^{th}$ best bid/ask quote instantaneously after the $t^{th}$ event. The volume of trade is denoted by $V^{mo}_t$ ($V^{mo}_t = 0$ when there is no trade) and the price of trade is denoted by $P^{mo}_t$ ($P^{mo}_t = 0$ when there is no trade, $P^{mo}_t = P_{t}^{b, 1}$ when a market order touches bid side and $P^{mo}_t = P_{t}^{a, 1}$ when a market order touches ask side). Moreover, we introduce six dummy variables $BLO_t$, $ALO_t$, $BMO_t$, $AMO_t$, $BTT_t$ and $ATT_t$ to indicate the direction of each event : bid side or ask side, respectively for limit order event ($BLO_t$ and $ALO_t$), market order event ($BMO_t$ and $AMO_t$) and trade-through event ($BTT_t$ and $ATT_t$). The definition of variables is detailed in Table ~\ref{Def_Variable}.

In order to capture the high-frequency dynamics in quotes and depths, we define a $K$-dimensional vector $$\mathbf{R}^1_t = [G_t^{b, L-1}, \cdots, G_t^{b, 1}, S_t, G_t^{a, 1}, \cdots, G_t^{a, L-1}, V_t^{b, L}, \cdots, V_t^{b, 1}, V_t^{a, 1}, \cdots, V_t^{a, L}]\;.$$
Modelling log prices and log volumes instead of absolute values is suggested by Potters and Bouchaud (2003) {~\cite{PB2003}} studying the statistical properties of market impacts and trades and can be found in many other empirical studies. Price and volume changes in log is interpreted as related changes in percentage.

Another vector of variables is denoted by $$\mathbf{R}^2_t = [BMO_t, AMO_t, BLO_t, ALO_t, BTT_t, ATT_t]\;,$$ indicating the nature of the $t^{th}$ event.

Table {~\ref{Tab_Stat}} provides a descriptive
statistics of the data used in this paper. It comprises limit order events, market order events and inter-trade price jump events. The analysis is done on two separated datasets: morning dataset (between $09h05$ and $13h15$) and afternoon dataset (between $13h15$ and $17h25$). We observe that there are more market order events in the afternoon than in the morning. Similarly, \emph{inter-trade price jump} events are slightly more frequent in the afternoon than in the morning. However, \emph{trade-through} events are more frequent in the morning than in the afternoon.

\begin{table}
\caption{Variable definitions}
\footnotesize
\label{Def_Variable}
\center
\begin{tabular}{cc}
\hline
Variable & Description \\
\hline
$P_t^{b, i}$ & the $i^{th}$ best log bid price just after the $t^{th}$ event \\
$P_t^{a, i}$ & the $i^{th}$ best log ask price just after the $t^{th}$ event \\
$G_t^{b, i}$ & the $i^{th}$ bid gap price just after the $t^{th}$ event \\
$S_t$ & the spread just after the $t^{th}$ event \\
$G_t^{a, i}$ & the $i^{th}$ ask gap price just after the $t^{th}$ event \\
$V_t^{b, i}$ & log volume of the $i^{th}$ best bid quote just after the $t^{th}$ event \\
$V_t^{a, i}$ & log volume of the $i^{th}$ best ask quote just after the $t^{th}$ event \\
$BLO_t$ & dummy variable equal to 1 if the $t^{th}$ event is a limit order event at bid side \\
$ALO_t$ & dummy variable equal to 1 if the $t^{th}$ event is a limit order event at ask side \\
$BMO_t$ & dummy variable equal to 1 if the $t^{th}$ event is a market order event at bid side \\
$AMO_t$ & dummy variable equal to 1 if the $t^{th}$ event is a market order event at ask side \\
$BTT_t$ & dummy variable equal to 1 if the $t^{th}$ event is a trade-through event at bid side \\
$ATT_t$ & dummy variable equal to 1 if the $t^{th}$ event is a trade-through event at ask side \\ \hline
\end{tabular}
\end{table}

\begin{table}
\scriptsize
\caption{Summary of limit order events, market order events and inter-trade price jump events, CAC40 stocks, April, 2011.}
\center
\begin{tabular}{ccccccccccccc}
\hline
\multirow{2}{*}{Stock} & \multicolumn{2}{c}{\#LO} & \multicolumn{2}{c}{\#MO} & \multicolumn{2}{c}{\#BidJump} & \multicolumn{2}{c}{\#AskJump} & \multicolumn{2}{c}{\#BidTT} & \multicolumn{2}{c}{\#AskTT}\\ \cline{2-13}
 & AM & PM & AM & PM & AM & PM & AM & PM & AM & PM & AM & PM \\ \hline
ACCP.PA	&	40505	&	44788	&	906	&	1067	 &	120	&	125	&	121	&	125	&	35	&	31	&	 37	&	36	\\
AIRP.PA	&	65775	&	83199	&	1358	&	1715	 &	190	&	257	&	201	&	264	&	55	&	66	&	 49	&	55	\\
ALSO.PA	&	92410	&	102069	&	1319	&	1590	 &	165	&	199	&	177	&	201	&	48	&	50	&	 43	&	39	\\
ALUA.PA	&	141048	&	110173	&	1900	&	2379	 &	202	&	233	&	214	&	242	&	84	&	96	&	 84	&	105	\\
AXAF.PA	&	156049	&	131722	&	1694	&	1951	 &	155	&	167	&	160	&	158	&	31	&	29	&	 35	&	26	\\
BNPP.PA	&	432091	&	236054	&	3164	&	3160	 &	377	&	368	&	386	&	372	&	104	&	77	&	 108	&	74	\\
BOUY.PA	&	28864	&	36112	&	973	&	1227	 &	103	&	140	&	114	&	154	&	29	&	33	&	 31	&	30	\\
CAGR.PA	&	133449	&	86078	&	1795	&	1645	 &	177	&	129	&	178	&	124	&	39	&	22	&	 45	&	23	\\
CAPP.PA	&	27679	&	30846	&	993	&	1262	 &	125	&	164	&	115	&	142	&	40	&	41	&	 37	&	40	\\
CARR.PA	&	111559	&	104513	&	1536	&	1799	 &	186	&	244	&	171	&	236	&	45	&	38	&	 53	&	43	\\
CNAT.PA	&	25216	&	29859	&	1144	&	1228	 &	142	&	147	&	125	&	126	&	39	&	31	&	 45	&	32	\\
DANO.PA	&	124929	&	106412	&	1618	&	2160	 &	188	&	260	&	188	&	267	&	57	&	60	&	 50	&	48	\\
EAD.PA	&	38720	&	35618	&	772	&	983	&	 98	&	102	&	72	&	87	&	23	&	23	&	18	 &	19	\\
EDF.PA	&	151715	&	75212	&	1881	&	1750	 &	190	&	158	&	182	&	153	&	62	&	33	&	 66	&	38	\\
ESSI.PA	&	21678	&	31743	&	528	&	751	&	 55	&	66	&	55	&	70	&	13	&	15	&	10	 &	9	\\
FTE.PA	&	78140	&	83328	&	1370	&	1710	 &	82	&	85	&	67	&	77	&	16	&	12	&	 14	&	12	\\
GSZ.PA	&	145293	&	105185	&	1781	&	2052	 &	190	&	217	&	160	&	189	&	48	&	43	&	 37	&	34	\\
ISPA.AS	&	165149	&	170835	&	1877	&	2663	 &	205	&	294	&	216	&	288	&	51	&	76	&	 61	&	72	\\
LAFP.PA	&	116640	&	87808	&	1149	&	1464	 &	167	&	213	&	177	&	217	&	61	&	54	&	 54	&	52	\\
LVMH.PA	&	80949	&	84063	&	1006	&	1181	 &	74	&	63	&	78	&	62	&	15	&	12	&	 15	&	12	\\
MICP.PA	&	101712	&	65052	&	1483	&	1564	 &	200	&	182	&	197	&	189	&	68	&	46	&	 55	&	45	\\
OREP.PA	&	33981	&	40395	&	1182	&	1393	 &	152	&	187	&	154	&	193	&	44	&	40	&	 42	&	39	\\
PERP.PA	&	65502	&	43922	&	971	&	1154	 &	133	&	124	&	136	&	141	&	31	&	22	&	 30	&	22	\\
PEUP.PA	&	51684	&	57536	&	1166	&	1258	 &	137	&	135	&	133	&	132	&	44	&	30	&	 43	&	33	\\
PRTP.PA	&	29682	&	31908	&	539	&	627	&	 36	&	34	&	44	&	36	&	6	&	6	&	5	 &	5	\\
PUBP.PA	&	71049	&	52461	&	1093	&	1350	 &	136	&	144	&	134	&	144	&	36	&	35	&	 36	&	31	\\
RENA.PA	&	136579	&	96872	&	1766	&	1843	 &	242	&	213	&	262	&	234	&	86	&	59	&	 103	&	75	\\
SASY.PA	&	111349	&	94709	&	1709	&	2221	 &	153	&	170	&	143	&	184	&	37	&	39	&	 37	&	36	\\
SCHN.PA	&	100690	&	82453	&	1297	&	1397	 &	97	&	78	&	92	&	86	&	23	&	12	&	 23	&	16	\\
SEVI.PA	&	33050	&	29122	&	619	&	659	&	 72	&	57	&	68	&	55	&	17	&	12	&	16	 &	11	\\
SGEF.PA	&	94111	&	65542	&	1454	&	1624	 &	178	&	198	&	169	&	199	&	59	&	53	&	 50	&	38	\\
SGOB.PA	&	158051	&	148326	&	1931	&	2336	 &	240	&	308	&	245	&	311	&	75	&	83	&	 73	&	78	\\
SOGN.PA	&	285430	&	172865	&	3455	&	3317	 &	496	&	475	&	450	&	456	&	182	&	150	&	 183	&	137	\\
STM.PA	&	96566	&	94170	&	1148	&	1367	 &	172	&	213	&	168	&	212	&	59	&	52	&	 63	&	59	\\
TECF.PA	&	77319	&	70533	&	1276	&	1405	 &	163	&	172	&	161	&	169	&	51	&	42	&	 45	&	35	\\
TOTF.PA	&	287449	&	224132	&	2388	&	3228	 &	308	&	396	&	301	&	418	&	81	&	101	&	 80	&	89	\\
UNBP.PA	&	41956	&	40961	&	612	&	755	&	 63	&	68	&	61	&	63	&	8	&	9	&	8	 &	8	\\
VIE.PA	&	49086	&	52755	&	936	&	1074	 &	106	&	105	&	108	&	104	&	19	&	15	&	 22	&	20	\\
VIV.PA	&	70356	&	73725	&	1478	&	1822	 &	130	&	151	&	122	&	143	&	31	&	24	&	 31	&	22	\\
VLLP.PA	&	45336	&	37801	&	1198	&	1471	 &	167	&	208	&	172	&	195	&	58	&	62	&	 54	&	50	\\ \hline
\end{tabular}
\label{Tab_Stat}
\end{table}

\section{Empirical facts : Bid-Ask liquidity balance and trade sign}
\label{section: EmpiricalFacts}

Before making an analysis on price jump prediction, we try to reveal whether the limit order volume information plays a role in determining the future market order's direction (trade sign). In order to study the conditional probability given the knowledge about bid/ask limit order liquidity, we propose a Bid-Ask volume ratio corresponding to depth $i$ just before the $k^{th}$ trade, which is defined as $W_{t_k - 1}(i)$ ($i \in \{1, \dots, L\}$), more precisely,
\begin{align}
W_{t_k - 1}(i) & = \log\left(\frac{\sum_{j=1}^i \exp(V^{b,j}_{t_k - 1})}{\sum_{j=1}^i \exp(V^{a, j}_{t_k - 1})}\right) \nonumber \\
& = \log\left(\sum_{j=1}^i \exp(V^{b,j}_{t_k - 1})\right) - \log\left(\sum_{j=1}^i \exp(V^{a, j}_{t_k - 1})\right) \;,
\end{align}
where $t_k$ is time index of the $k^{th}$ market order event.

For all $x \in \mathbb{R}_+$, the conditional probability of a future buy market order (positive trade sign) that the next trade is triggered by a buy market order given $V_{t_k}(i) \geq x$ is defined as,
\begin{align}
& \mathbb{P}\left(I^{ts}_{t_{k}}=1 | W_{t_k - 1}(i) \geq x \right) \;.
\end{align}
where we denote the TradeSign at time $t_k$ by $I^{ts}_{t_k}$.

Similarly, for all $x \in \mathbb{R}_+$, the conditional probability of a future sell market order (negative trade sign) that the next trade is triggered by a sell market order given $V_{t_k}(i) \geq x$ is defined as,

\begin{align}
& \mathbb{P}\left(I^{ts}_{t_k}=-1 | W_{t_k - 1}(i) \leq x \right) \;.
\end{align}

\begin{figure}[!h]
\caption{The conditional probability of a buy market order vs bid-ask volume ratio, April, 2011.}
\center
\includegraphics[trim=0mm 0mm 0mm 0mm, clip, height=6.0cm, width=6.0cm, angle=-90]{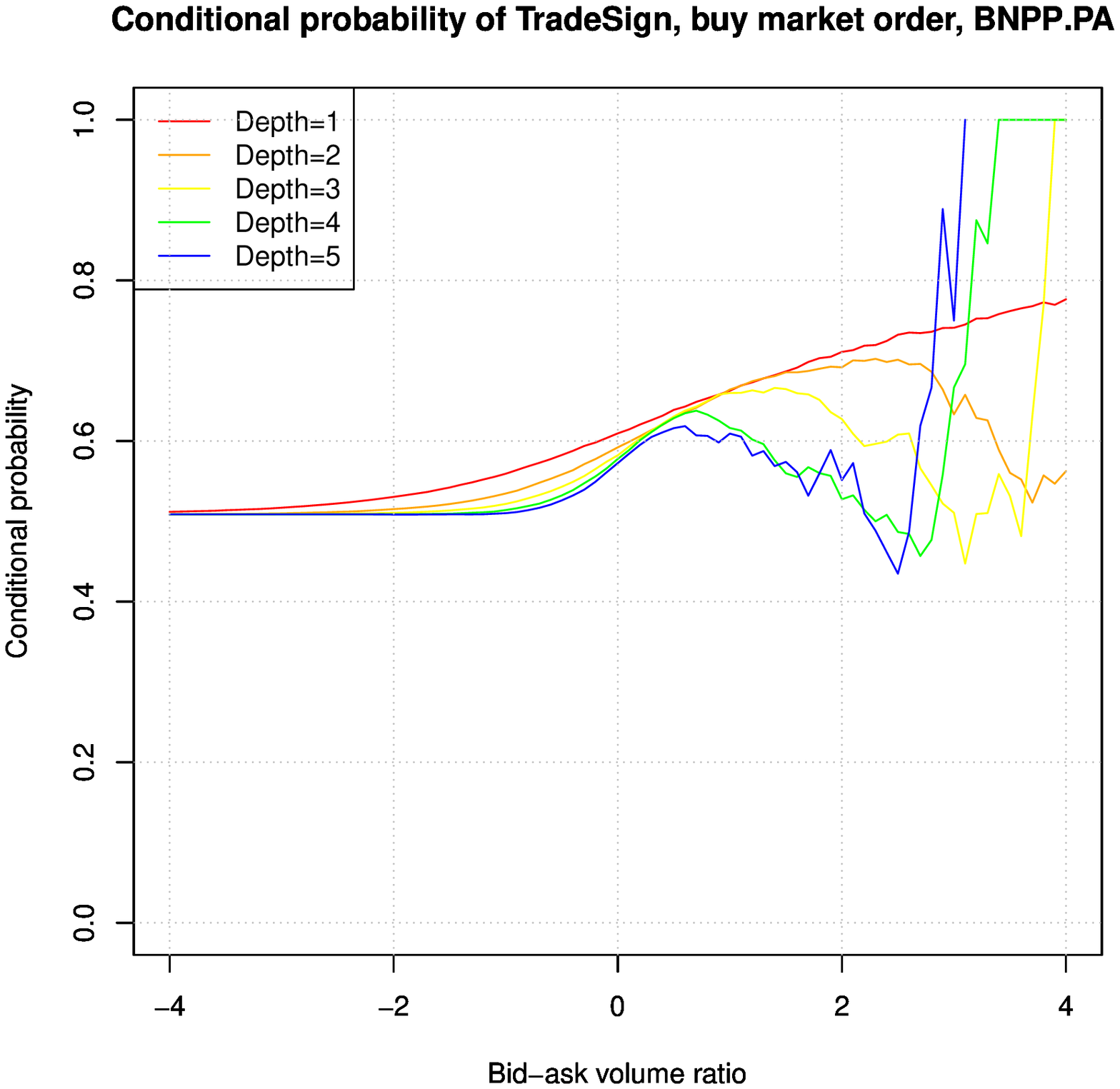}
\includegraphics[trim=0mm 0mm 0mm 0mm, clip, height=6.0cm, width=6.0cm, angle=-90]{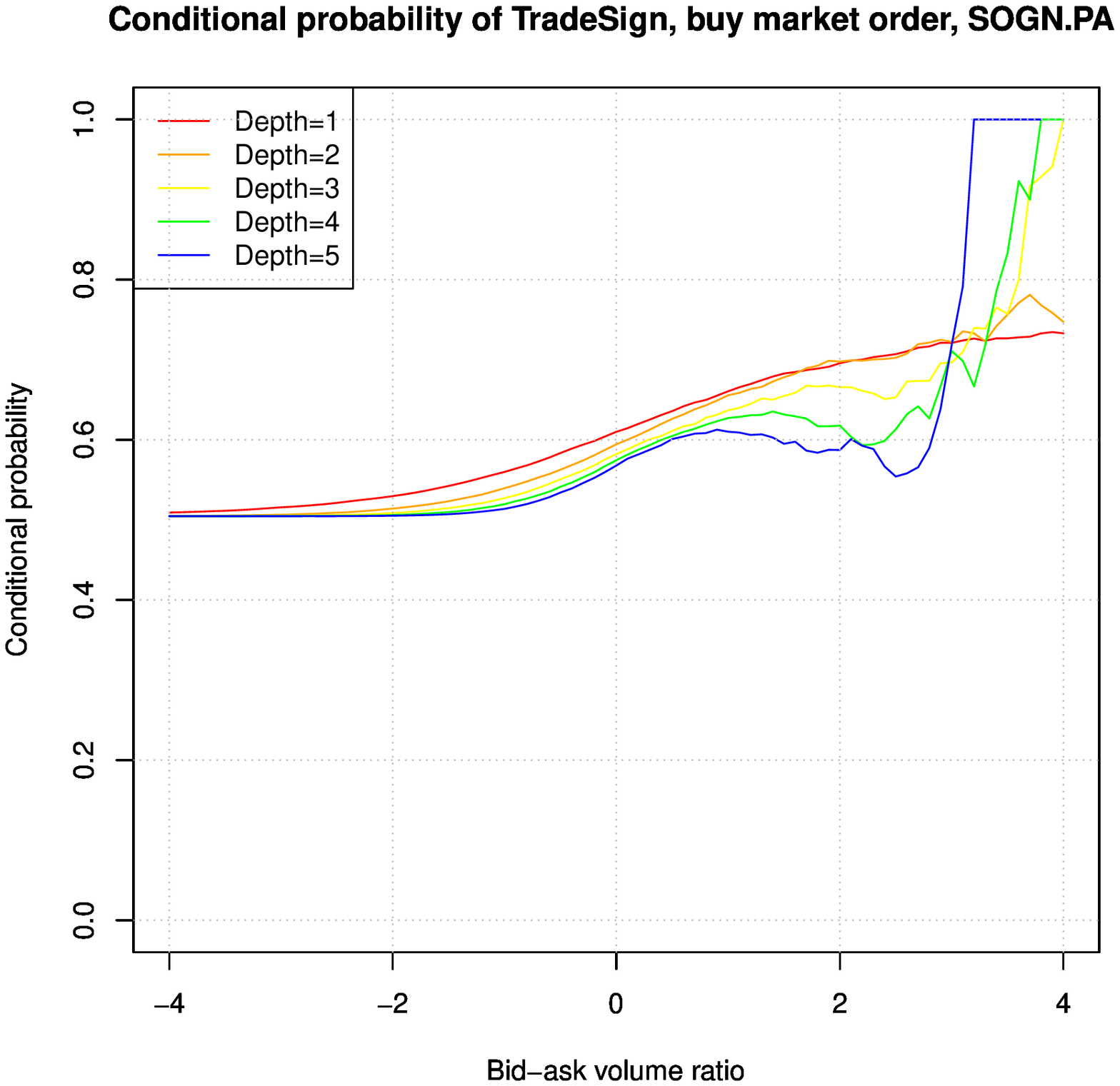} \\
\includegraphics[trim=0mm 0mm 0mm 0mm, clip, height=6.0cm, width=6.0cm, angle=-90]{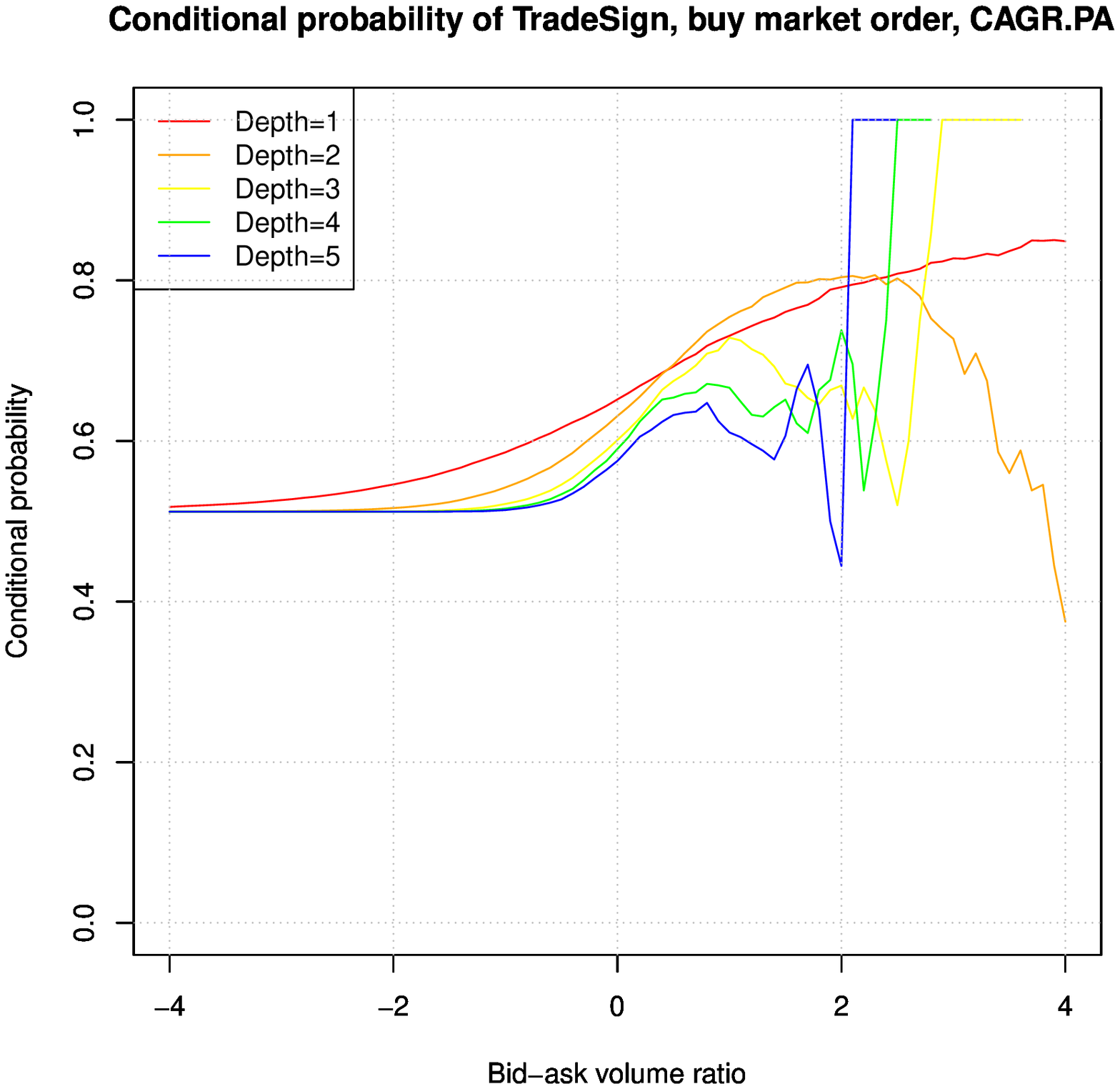}
\includegraphics[trim=0mm 0mm 0mm 0mm, clip, height=6.0cm, width=6.0cm, angle=-90]{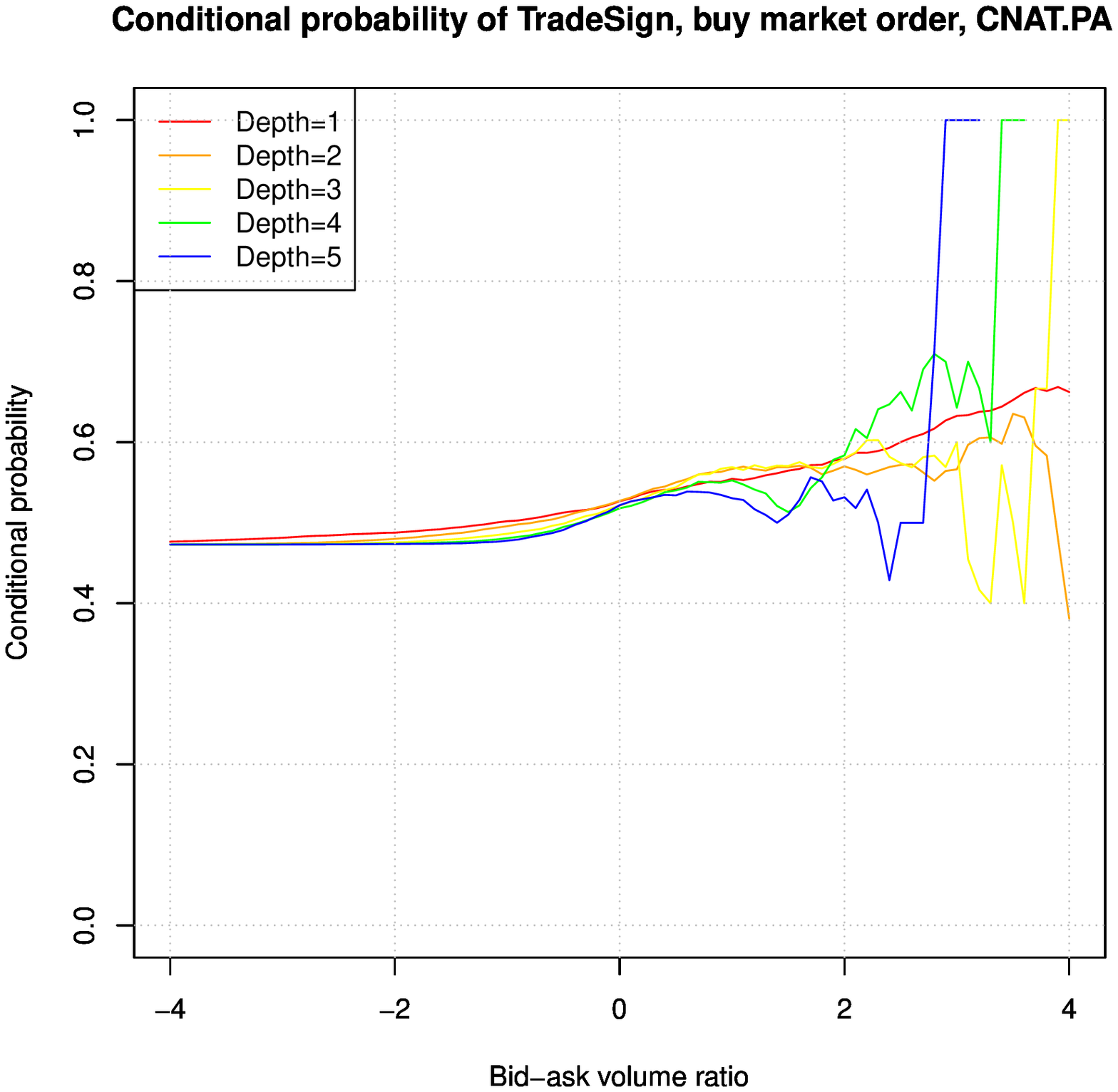} \\
\label{Figure_Vlm_BMO}
\end{figure}

\begin{figure}[!h]
\caption{The conditional probability of a sell market order vs bid-ask volume ratio, April, 2011.}
\center
\includegraphics[trim=0mm 0mm 0mm 0mm, clip, height=6.0cm, width=6.0cm, angle=-90]{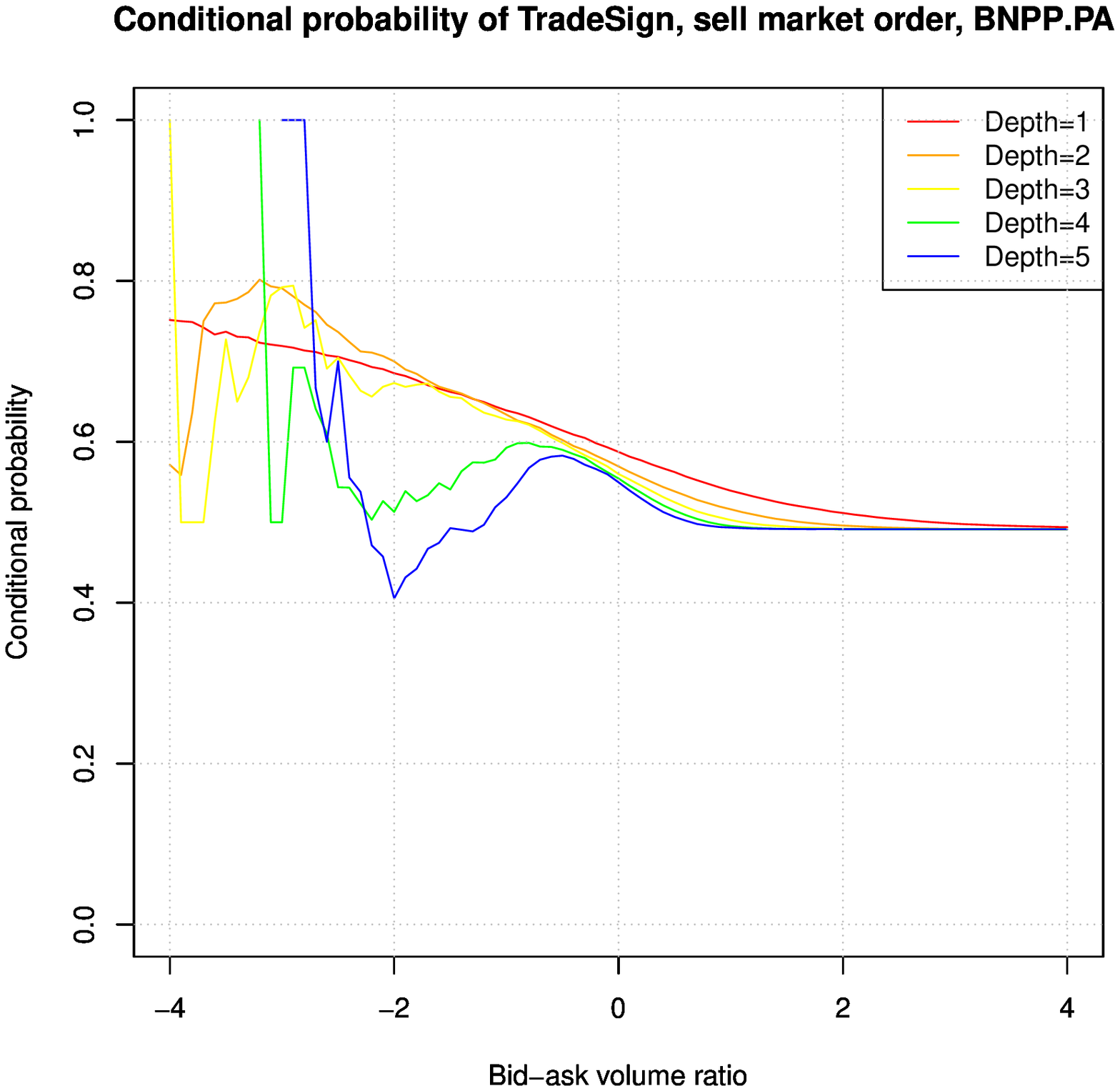}
\includegraphics[trim=0mm 0mm 0mm 0mm, clip, height=6.0cm, width=6.0cm, angle=-90]{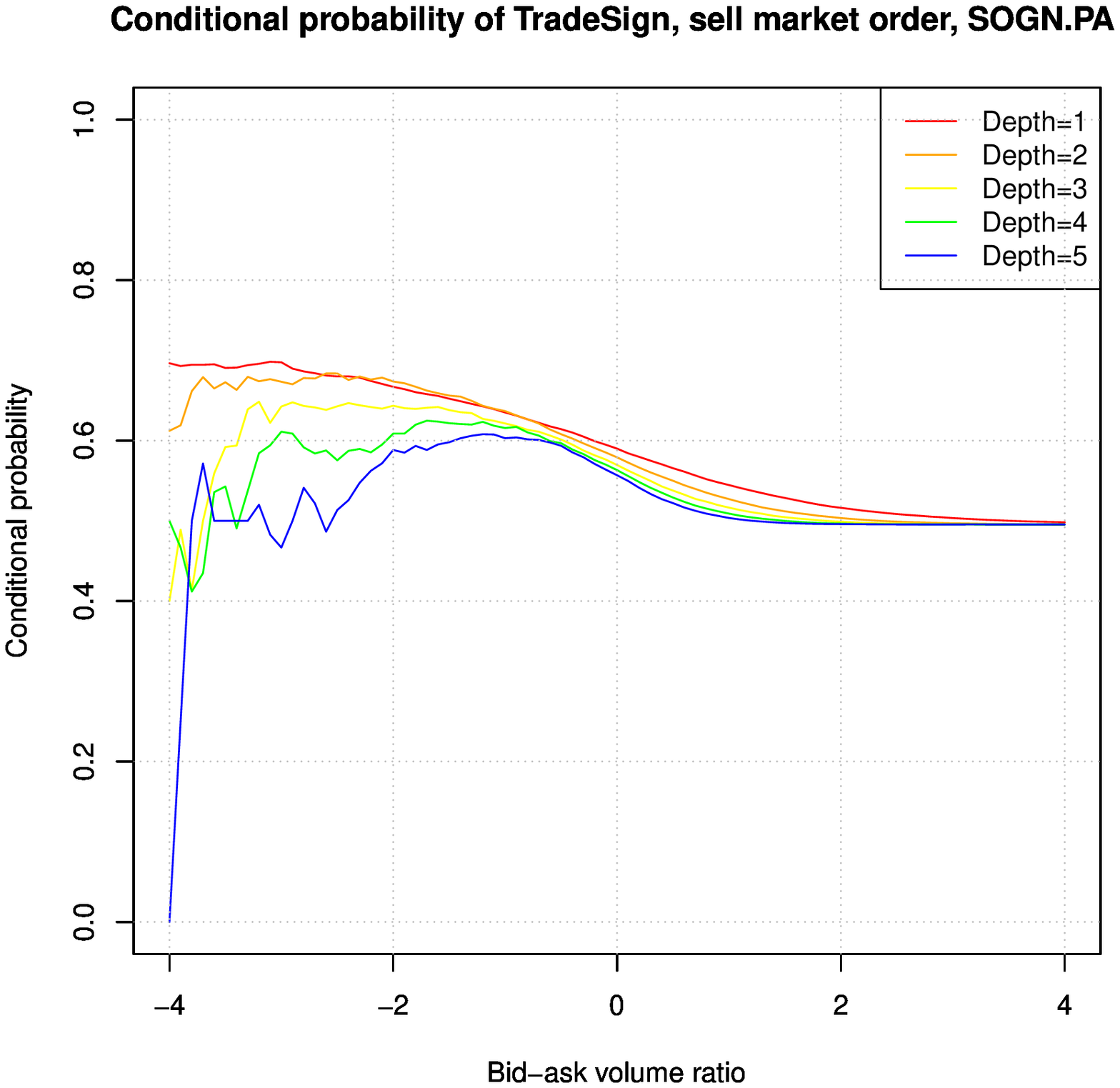} \\
\includegraphics[trim=0mm 0mm 0mm 0mm, clip, height=6.0cm, width=6.0cm, angle=-90]{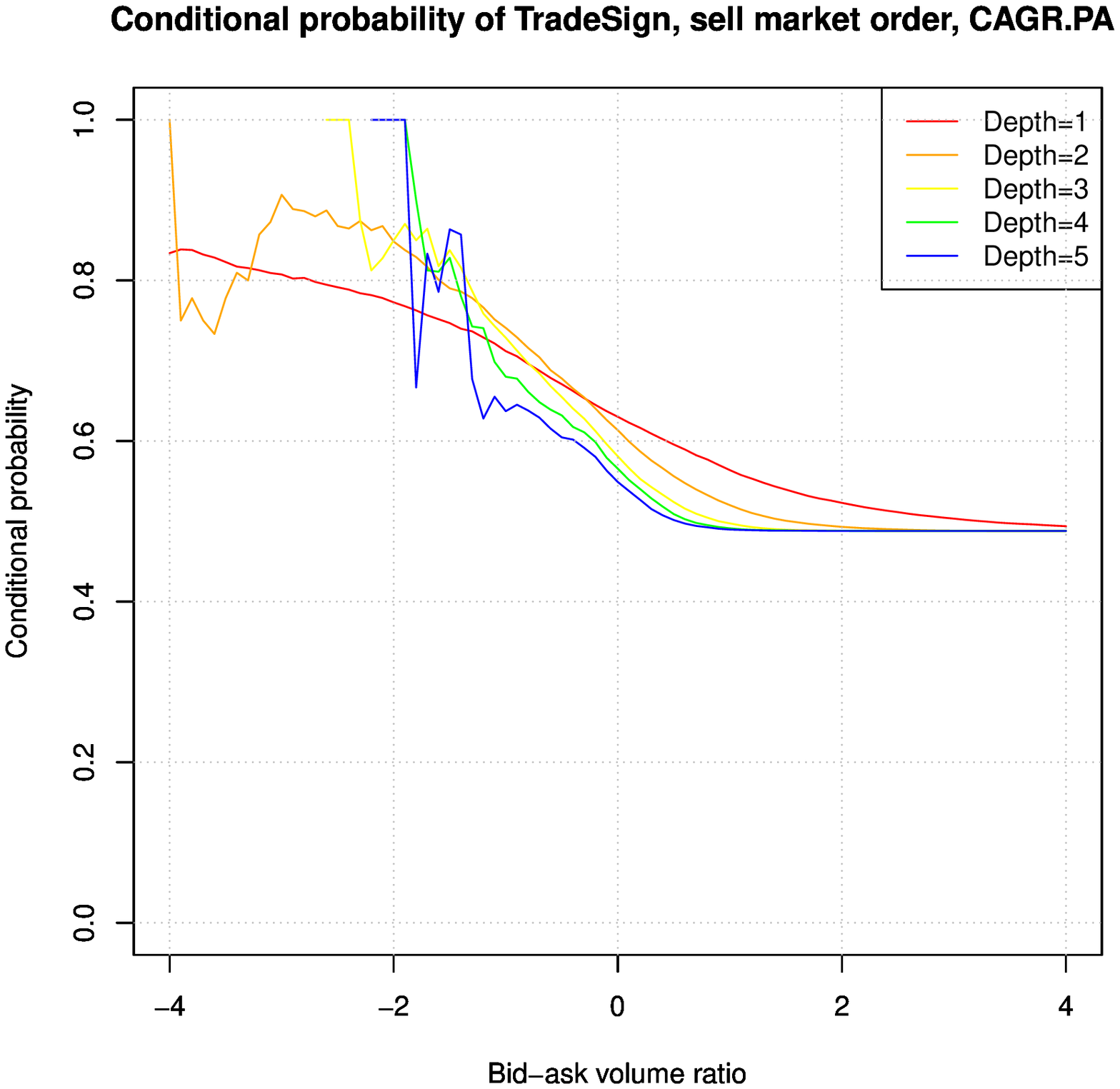}
\includegraphics[trim=0mm 0mm 0mm 0mm, clip, height=6.0cm, width=6.0cm, angle=-90]{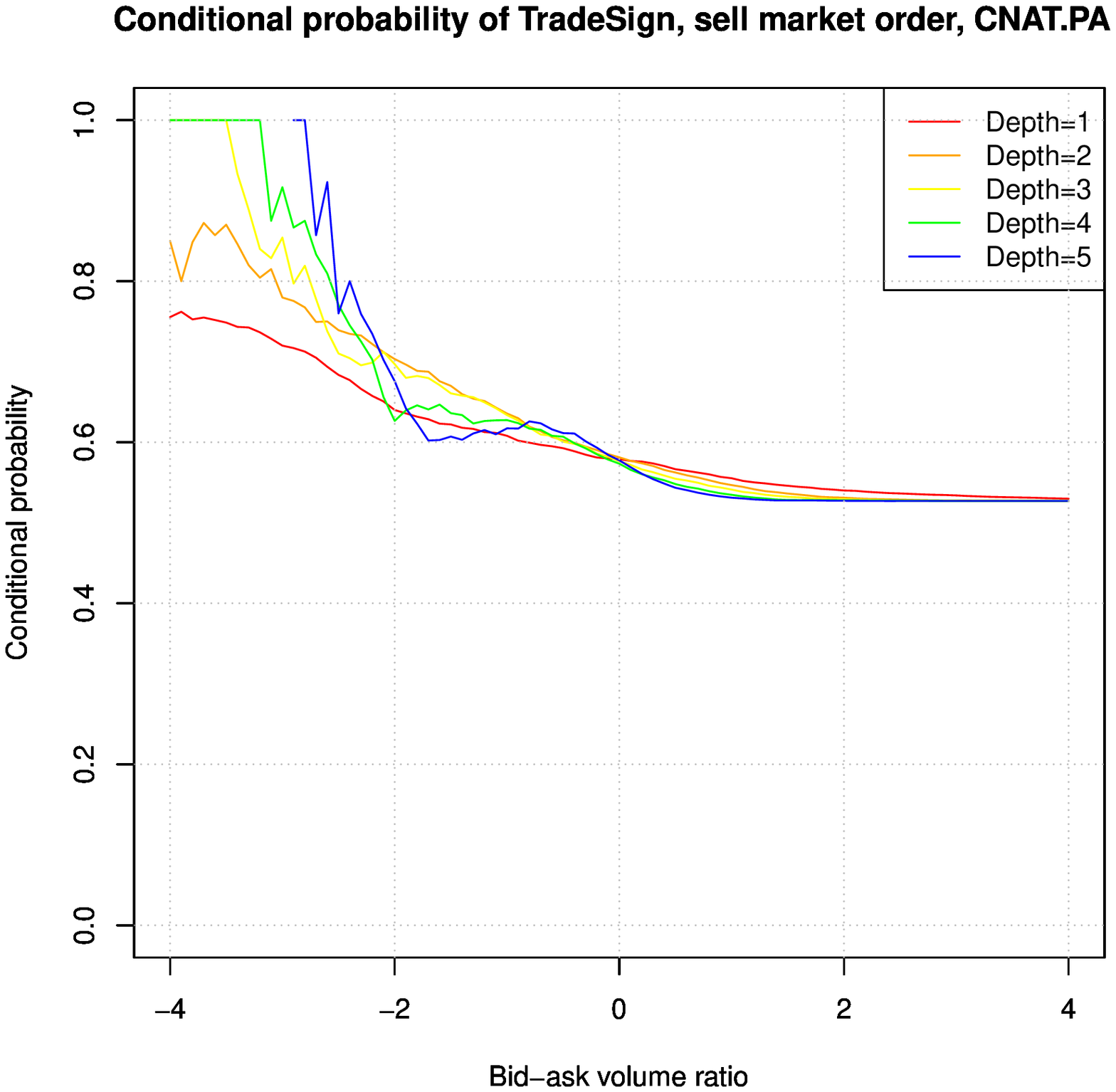} \\
\label{Figure_Vlm_SMO}
\end{figure}

The relationship between $\mathbb{P}\left(I^{ts}_{t_k}=1 | W_{t_k - 1}(i) \geq x \right)$ and $x$ for $i \in \{1, \dots, L\}$ is shown in Figure~\ref{Figure_Vlm_BMO} and~\ref{Figure_Vlm_SMO}. We observe that the conditional probability of the next trade sign is highly correlated with the Bid-Ask volume ratio corresponding to depth $1$. Nevertheless, the dependance between the conditional probability of the next trade sign and the Bid-Ask volume ratio corresponding to depth larger than $1$ is much more noised.  Figure~\ref{Figure_Vlm_BMO_CAC40} in Appendix shows the relationship between $\mathbb{P}\left(I^{ts}_{t_k}=1 | W_{t_k - 1}(1) \geq x \right)$ and $x$ for all stocks of CAC40. It is worth remarking that the trade sign's conditional probability reaches $0.80$ in average when the liquidity on the best limit prices is quite unbalanced.

\section{Logistic regression analysis and variable selection by LASSO}
\label{section: Logit}
\subsection{Logistic regression analysis}
The result shown in the previous section reveals that Bid-Ask liquidity balance provides important information on the incoming market order. In this section, we introduce the standard logistic regression to predict the \emph{inter-trade price jump} occurrence and use $LASSO$ select regularization to evidence the importance of each variable in this prediction.

We denote the number of market order events by $N$ and for each $i \in \{1, \cdots, N\}$, $\mathbf{X}_i=[1, V^{mo}_{t_i}, \mathbf{R}^1_{t_i}, \mathbf{R}^1_{t_i - 1}, \cdots, \mathbf{R}^1_{t_{i}-m+1}, \mathbf{R}^2_t, \mathbf{R}^2_{t_i-1}, \cdots, \mathbf{R}^2_{t_{i}-n+1}]$ ($X \in \mathbb{R}^{(p+1) \times 1}$, $p = m(2L-1) + 6n$) the explanatory variables summarizing the available order book information when the $t^{th}$ event is a market order event, $y_i$ is a binary variable indicating whether the event is an bid/ask \emph{inter-trade price jump}, $y_i$ is defined as follows,

\begin{align}
\text{Bid side inter-trade price jump indicator: } Y_i =
\begin{cases}
 1, & \text{if } P_{t_{i+1}}^{mo} < P_{t_i}^{b, 1} \\
 0, & \text{ otherwise}
\end{cases}
\end{align}

or

\begin{align}
\text{Ask side inter-trade price jump indicator: } Y_i =
\begin{cases}
 1, & \text{if } P_{t_{i+1}}^{mo} > P_{t_i}^{a, 1} \\
 0, & \text{ otherwise}
\end{cases}
\end{align}

In the logistic model, the probability of the bid/ask \emph{inter-trade price jump} occurrence is assumed to be given by:
\begin{align}
\label{eq: Logit_CondProba}
\log{\frac{\mathrm{P_{\boldsymbol{\beta}}}(Y=1|\mathbf{X})}{1-\mathrm{P_{\boldsymbol{\beta}}}(Y=1|\mathbf{X})}}=\boldsymbol{\beta}^{T}\mathbf{X}\;,
\end{align}
where $\boldsymbol{\beta}=[\beta_0, \beta_1, \cdots, \beta_p]^T$.

Observing that for $i = 1$
\begin{align}
& W_{t_k-1}(i)= V^{b,1}_{t_k-1} - V^{a, 1}_{t_k-1} \;,
\end{align}
we see that the linearity of the conditional probability $\mathrm{P_{\boldsymbol{\beta}}}(Y=1|\mathbf{X})$ on variables $V^{b,1}_{t_k}$ and $V^{a, 1}_{t_k}$ in Equation~\ref{eq: Logit_CondProba} allows us to capture the contribution of $W_{t_k}(i)$ in the prediction.

The parameters $\boldsymbol{\beta}$ are unknown and should be estimated from the data. We use the maximum likelihood to estimate the parameters. It is well known that the log-likelihood function given by
\begin{equation}
\label{eq: LOGIT1}
\mathcal{L}(\boldsymbol{\beta})=\sum_{i=1}^{N}{\{\log(1+e^{\boldsymbol{\beta}^T \mathbf{X}_i})-y_i\boldsymbol{\beta}^T \mathbf{X}_i\}}\;.
\end{equation}

The likelihood function is convex and therefore can be optimized using a standard optimization method.

\subsection{Variable selection by LASSO}
Since the number of explanatory variables $p$ being quite large, it is of interest to perform a variable selection procedure to select the most important variables. A classical variable selection procedure when the number of regressors is large is the LASSO procedure see Hastie et al (2003) {~\cite{Hastie2003}}. Instead of using a $BIC$ penalization, the LASSO procedure adds to the likelihood the norm of the logistic coefficient, which is known to induce a sparse solution. This penalization induces an automatic variable selection effect.

The LASSO estimate for logistic regression is defined by
\begin{eqnarray}
\label{eq: LASSO2}
\hat{\beta}^{lasso}(\lambda) &=& \underset{\beta}{\mathrm{argmin}} \sum_{i=1}^{N}\left(\left(-\log(1+e^{\beta^T \mathbf{X}_i})+ Y_i\beta^T \mathbf{X}_i\right) + \lambda \sum_{j=1}^{p}{|\beta_j|}\right)\;.
\end{eqnarray}
The constraint on $\sum_{j=1}^{p}{|\beta_j|}$ makes the solutions nonlinear in the $y_i$ and there is no closed form expression as in ridge regression. Because of the nature of constraint, making $\lambda$ sufficiently large will cause some of the coefficients to be exactly zero. Germain et Roueff (2009) {~\cite{germain-roueff-2010}} gives the uniform consistency and a functional central limit theorem for the LASSO regularization path for the general linear model.

\subsection{Results}
Choosing $L = 5$, $m=5$ and $n=5$, the dimension of limit order book's profile vector is $p=1 + m(2L-1) + 6n = 76$.

The parameter $\lambda$ in LASSO is estimated by cross-validation, then we calculate AUC value (area under ROC curve) to measure the prediction quality. A ROC (receiver operating characteristic) curve is a graphical plot of the true positive rate vs. false positive rate. The area under the ROC curve is a good measure to measuring the model prediction quality. The AUC value is equal to the probability that a classifier will rank a randomly chosen positive instance higher than a randomly chosen negative one.

\begin{figure}[!h]
\caption{AUC value, price jump prediction, CAC40, April, 2011.}
\center
\includegraphics[trim=0mm 0mm 0mm 0mm, clip, height=6cm, width=6cm, angle=-90]
{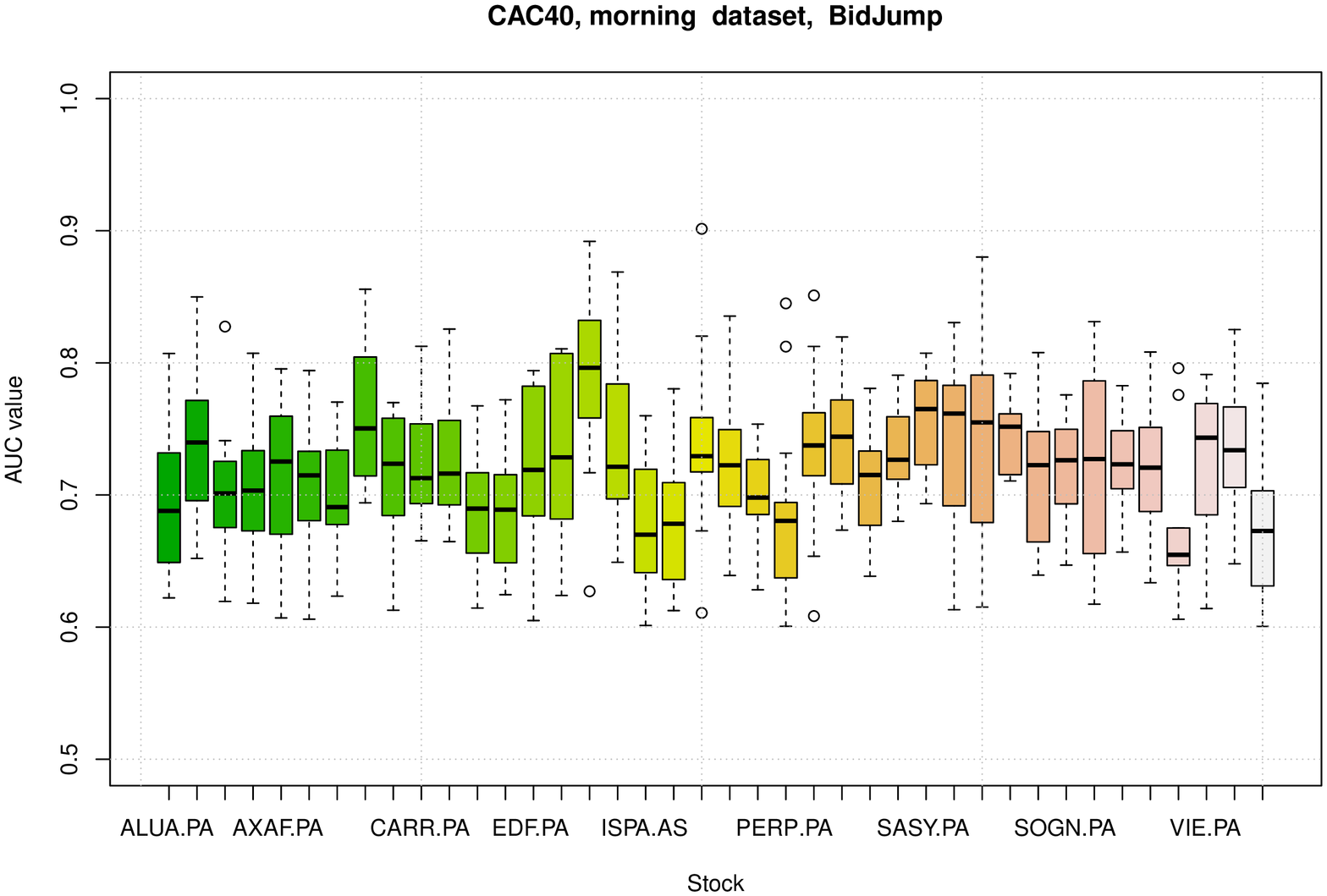}
\includegraphics[trim=0mm 0mm 0mm 0mm, clip, height=6cm, width=6cm, angle=-90]
{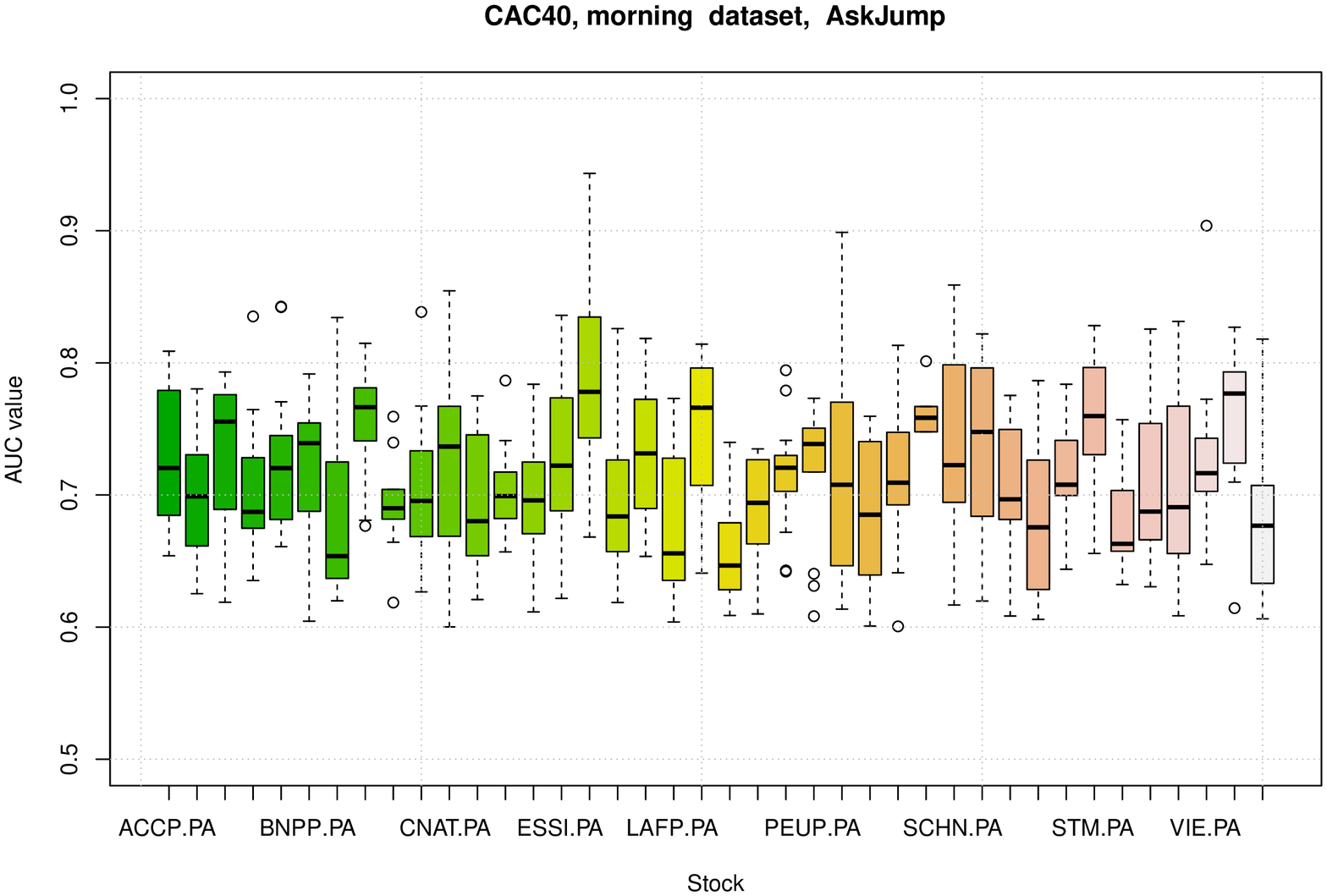} \\
\includegraphics[trim=0mm 0mm 0mm 0mm, clip, height=6cm, width=6cm, angle=-90]
{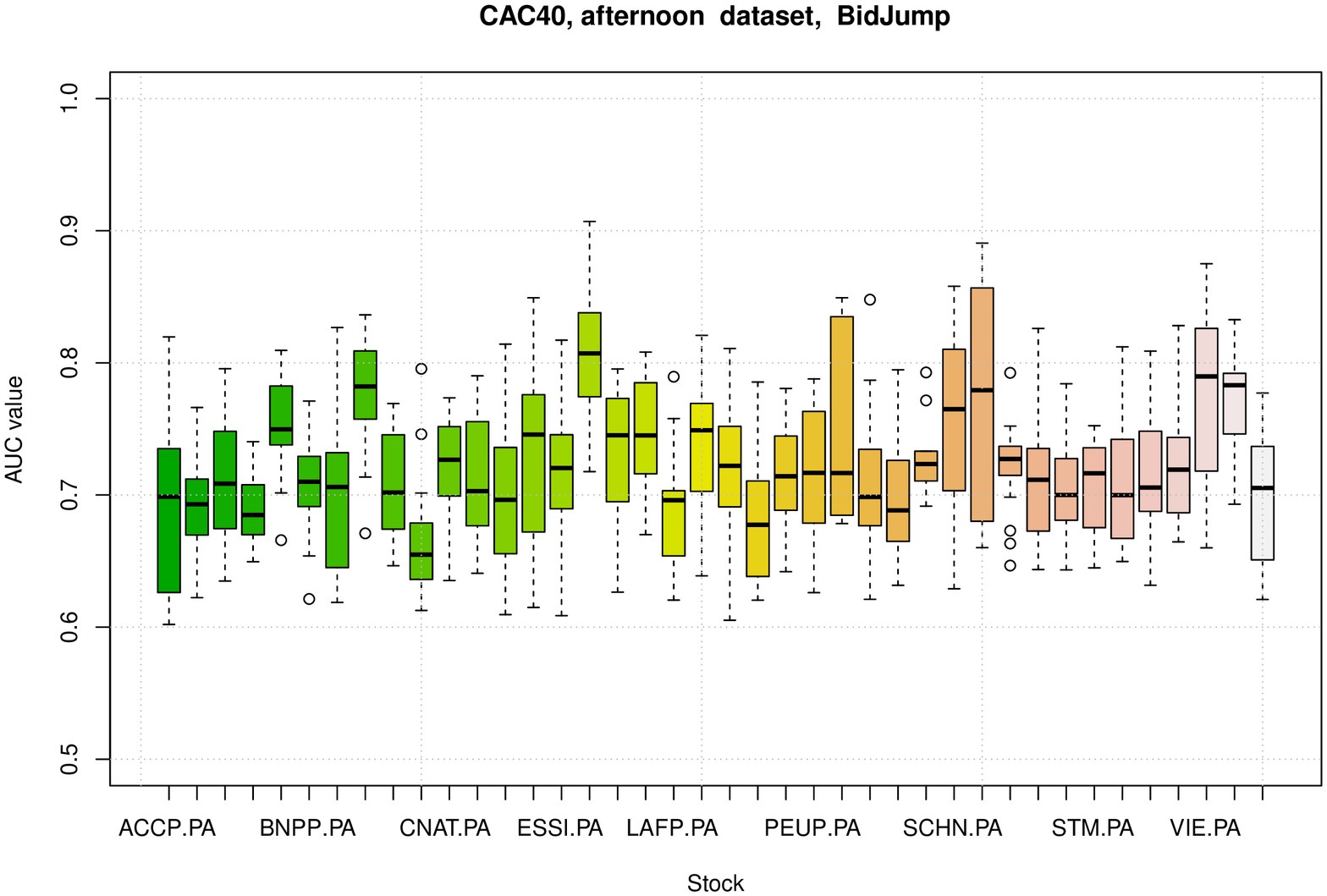}
\includegraphics[trim=0mm 0mm 0mm 0mm, clip, height=6cm, width=6cm, angle=-90]
{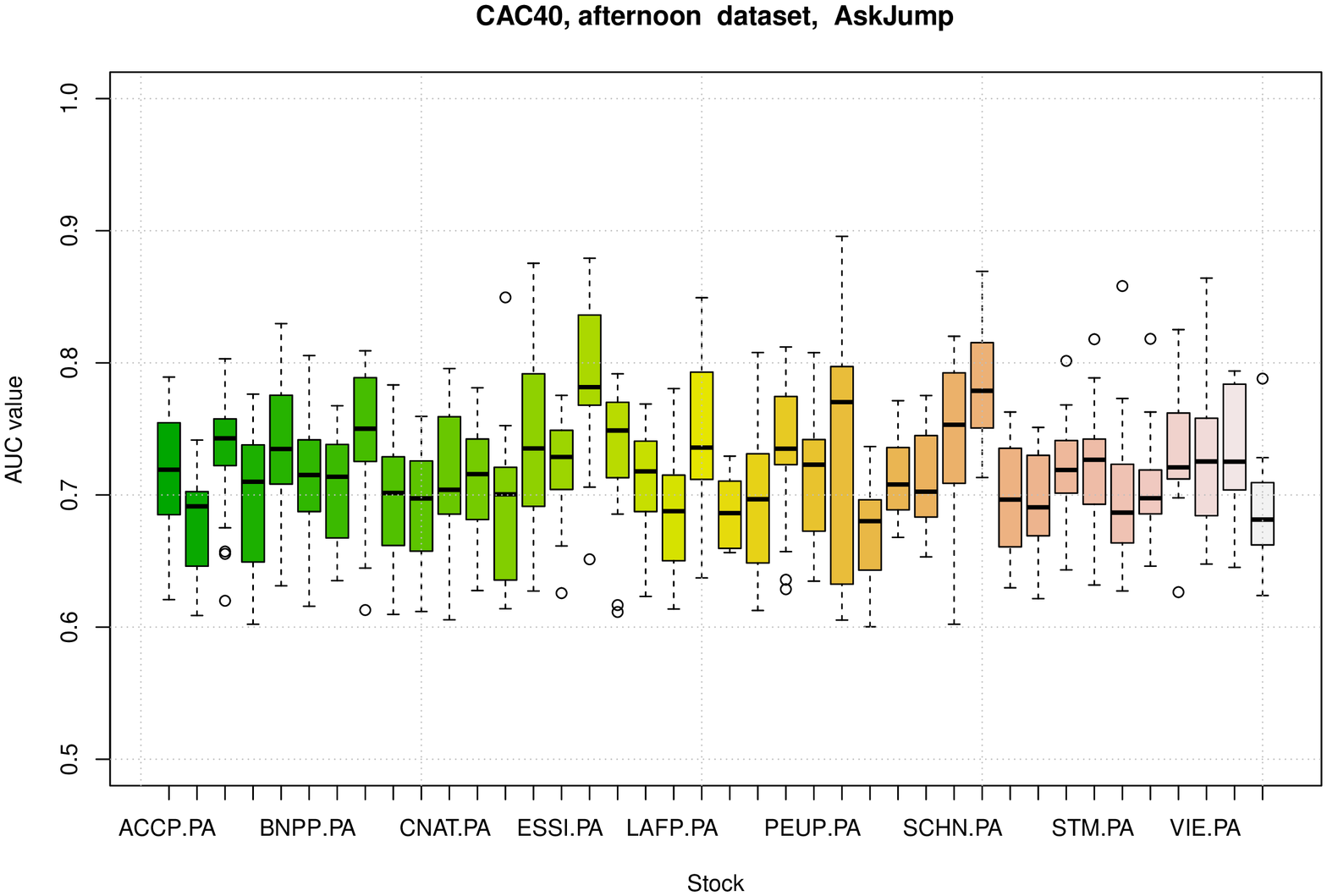}\\
\includegraphics[trim=0mm 0mm 0mm 0mm, clip, height=6cm, width=6cm, angle=-90]
{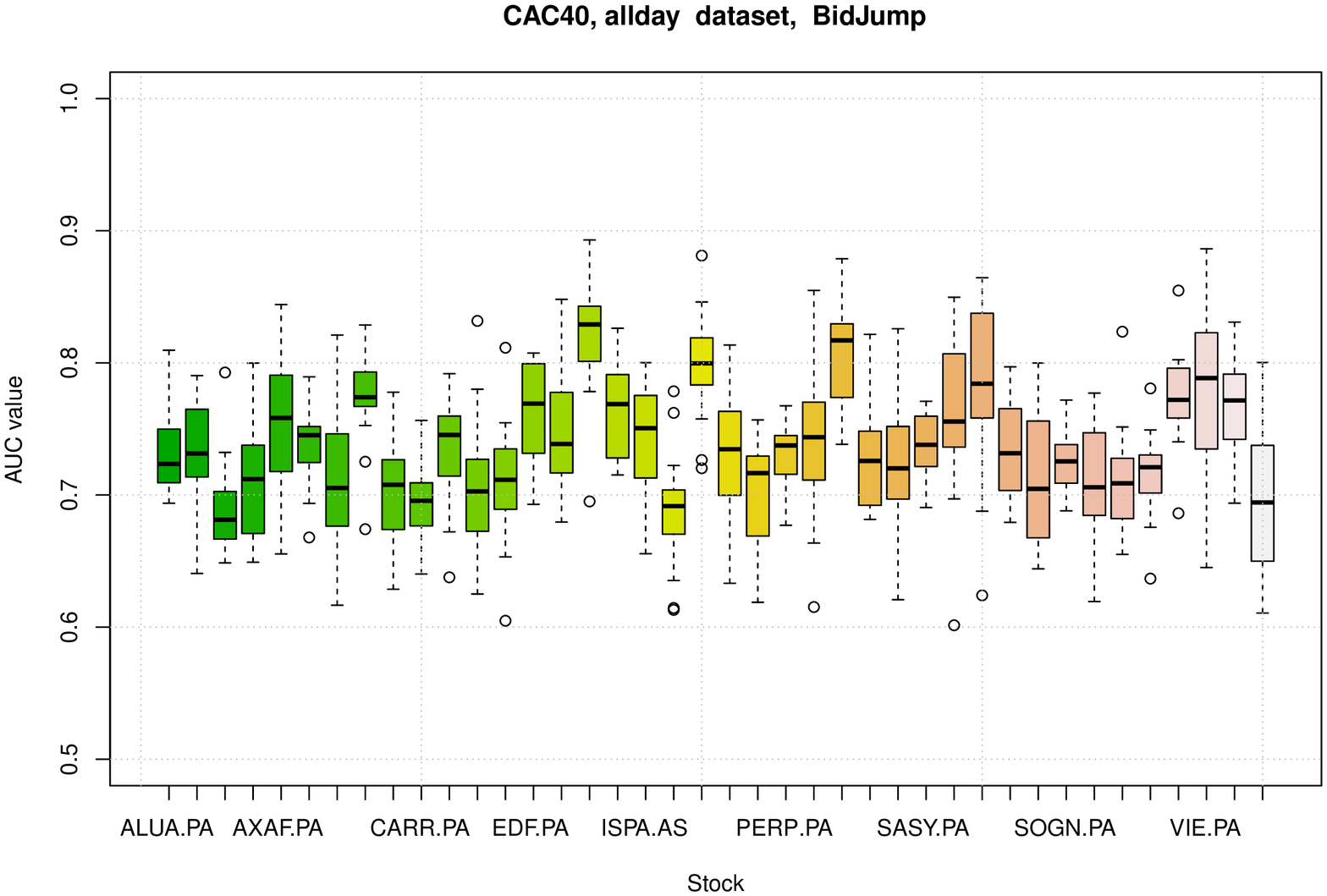}
\includegraphics[trim=0mm 0mm 0mm 0mm, clip, height=6cm, width=6cm, angle=-90]
{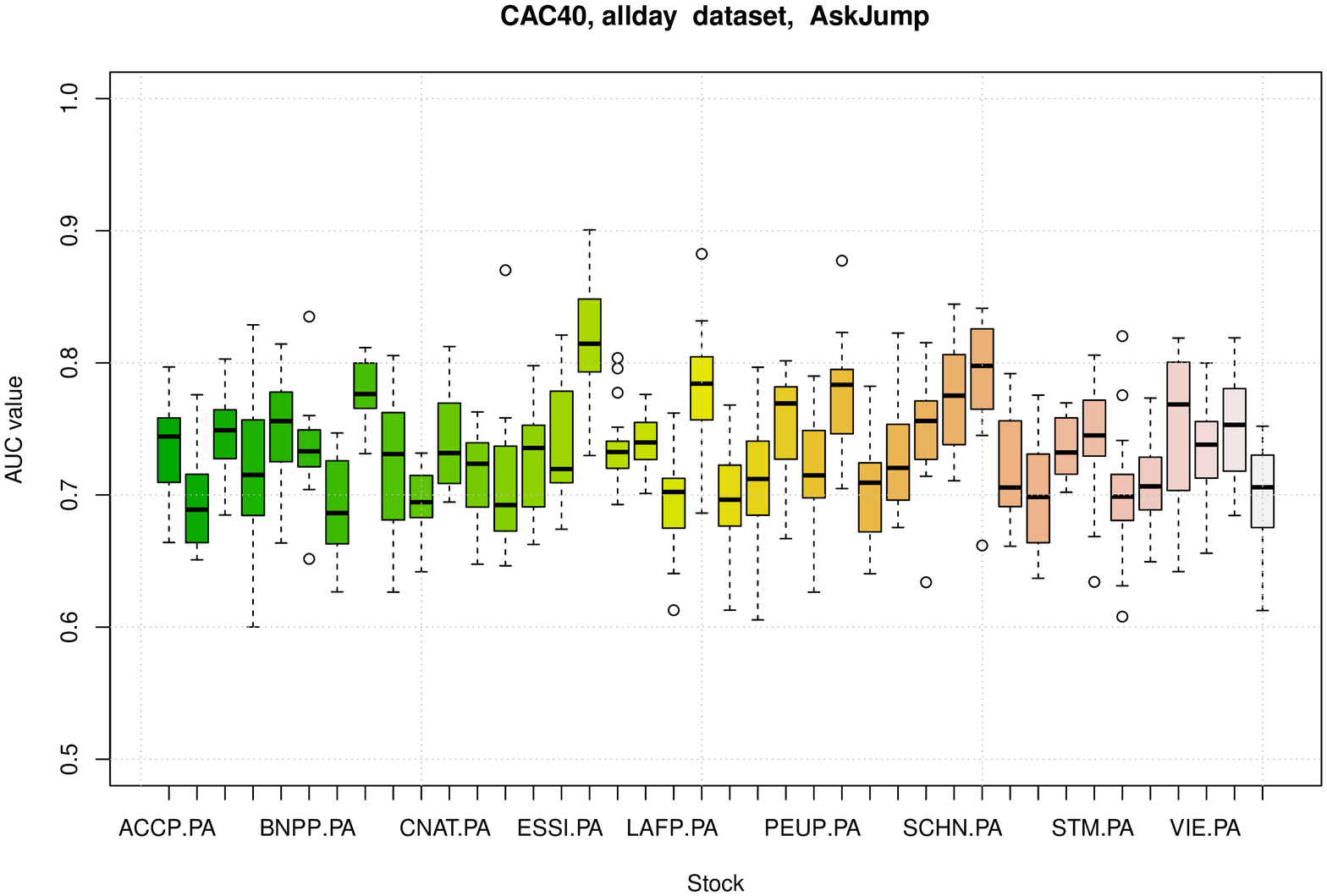} \\
\label{AUC_Jump}
\end{figure}

We show the out-of-sample AUC value in Figure ~\ref{AUC_Jump}. The stocks are sorted in alphabetic order. We see that for each prediction task, the AUC value is around $0.80$ and it is consistently high over all datasets and all stocks of $CAC40$.

% from here
In order to discover the contribution of each variable to the prediction, we add an analysis on the five firstly selected variables for each prediction task of all stocks of $CAC40$ (with allday dataset).

\begin{figure}[!h]
\caption{Variable selection for BidJump prediction, CAC40, April, 2011. From left to right, from top to bottom, each figure shows how many times a variable is selected as the $k^{th}$ selected variable by $LASSO$, $k=\{1, \dots, 5\}$.}
\center
\includegraphics[trim=0mm 0mm 0mm 0mm, clip, height=6cm, width=6cm, angle=-90]
{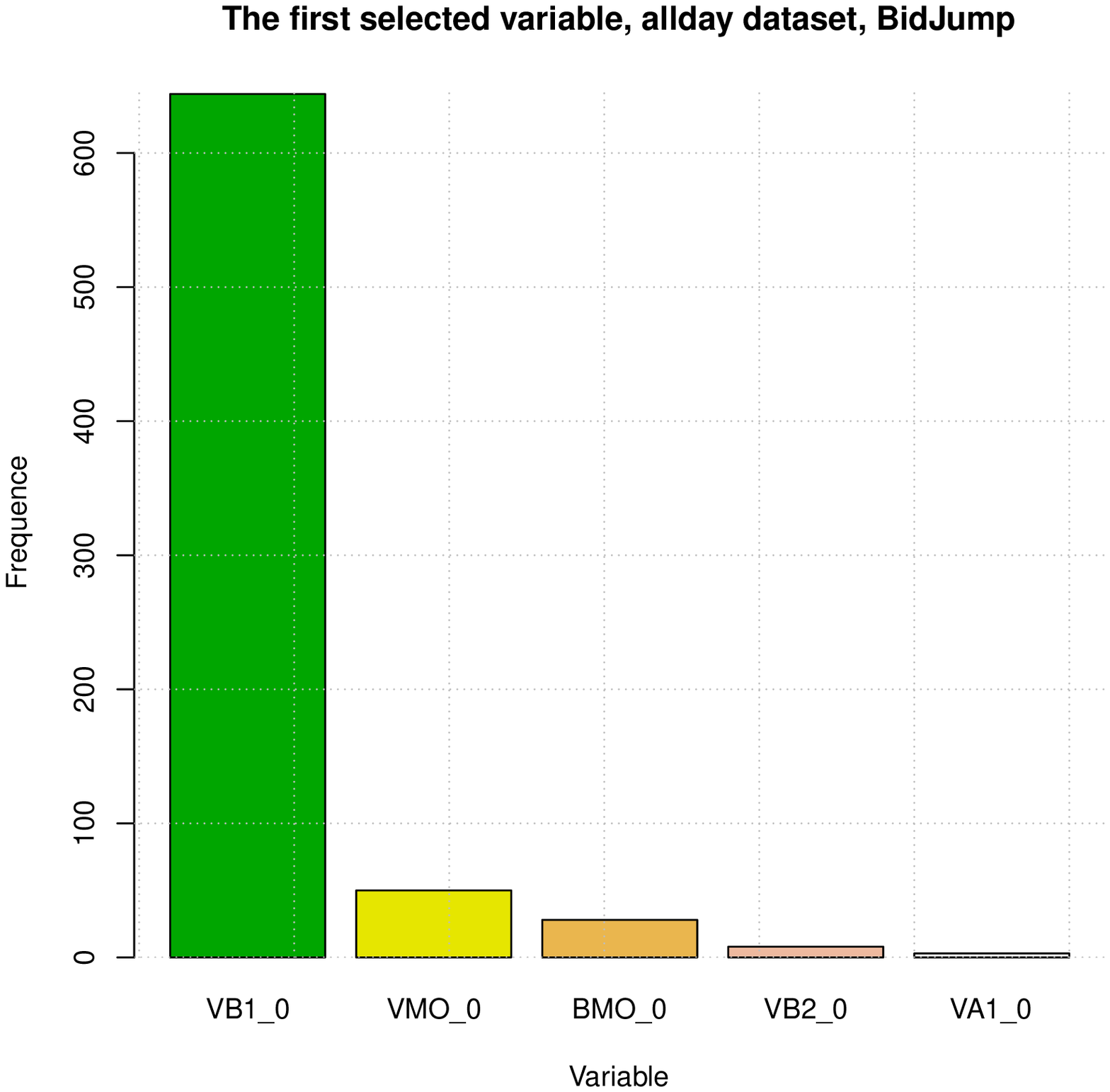}
\includegraphics[trim=0mm 0mm 0mm 0mm, clip, height=6cm, width=6cm, angle=-90]
{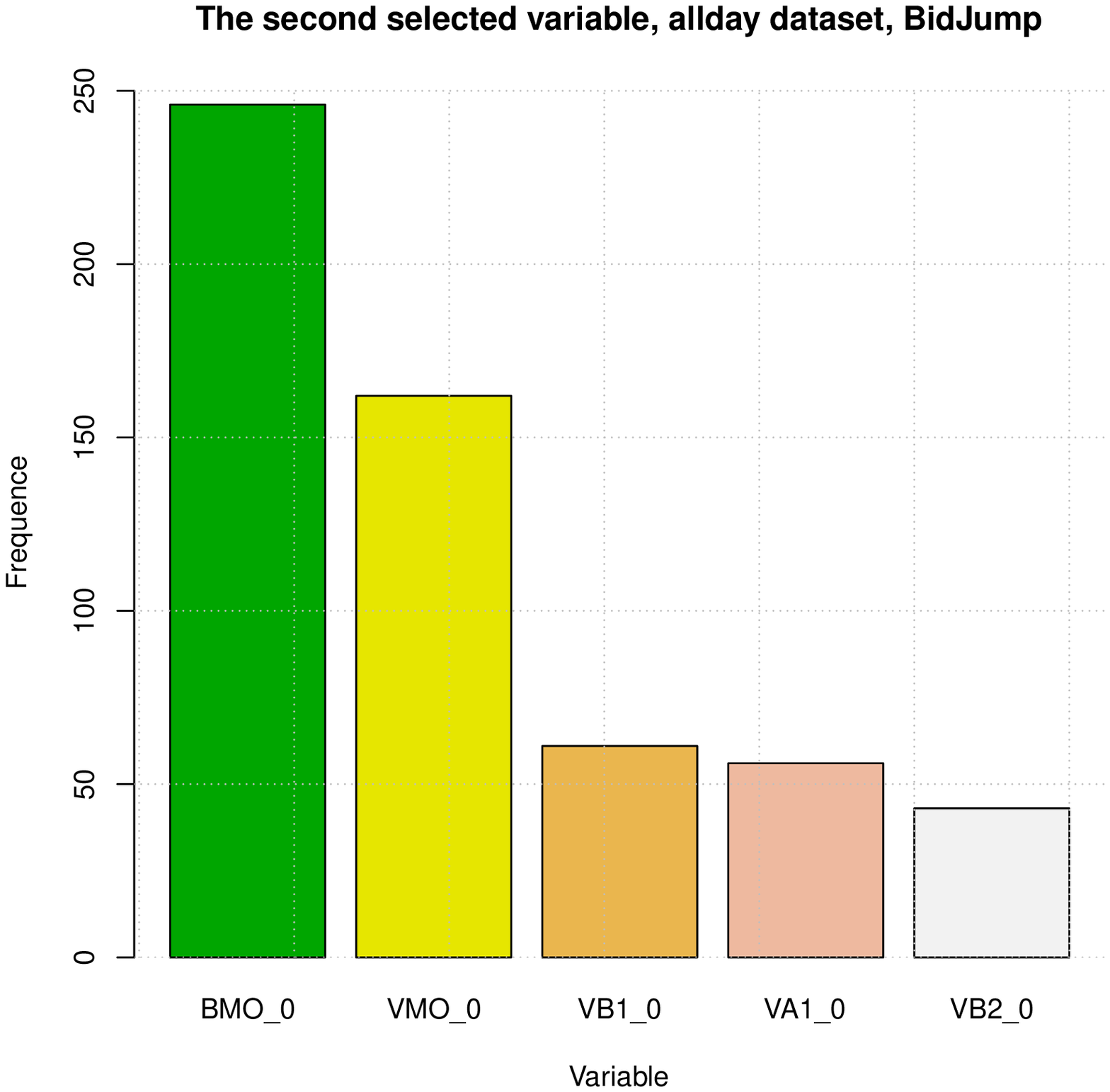}\\
\includegraphics[trim=0mm 0mm 0mm 0mm, clip, height=6cm, width=6cm, angle=-90]
{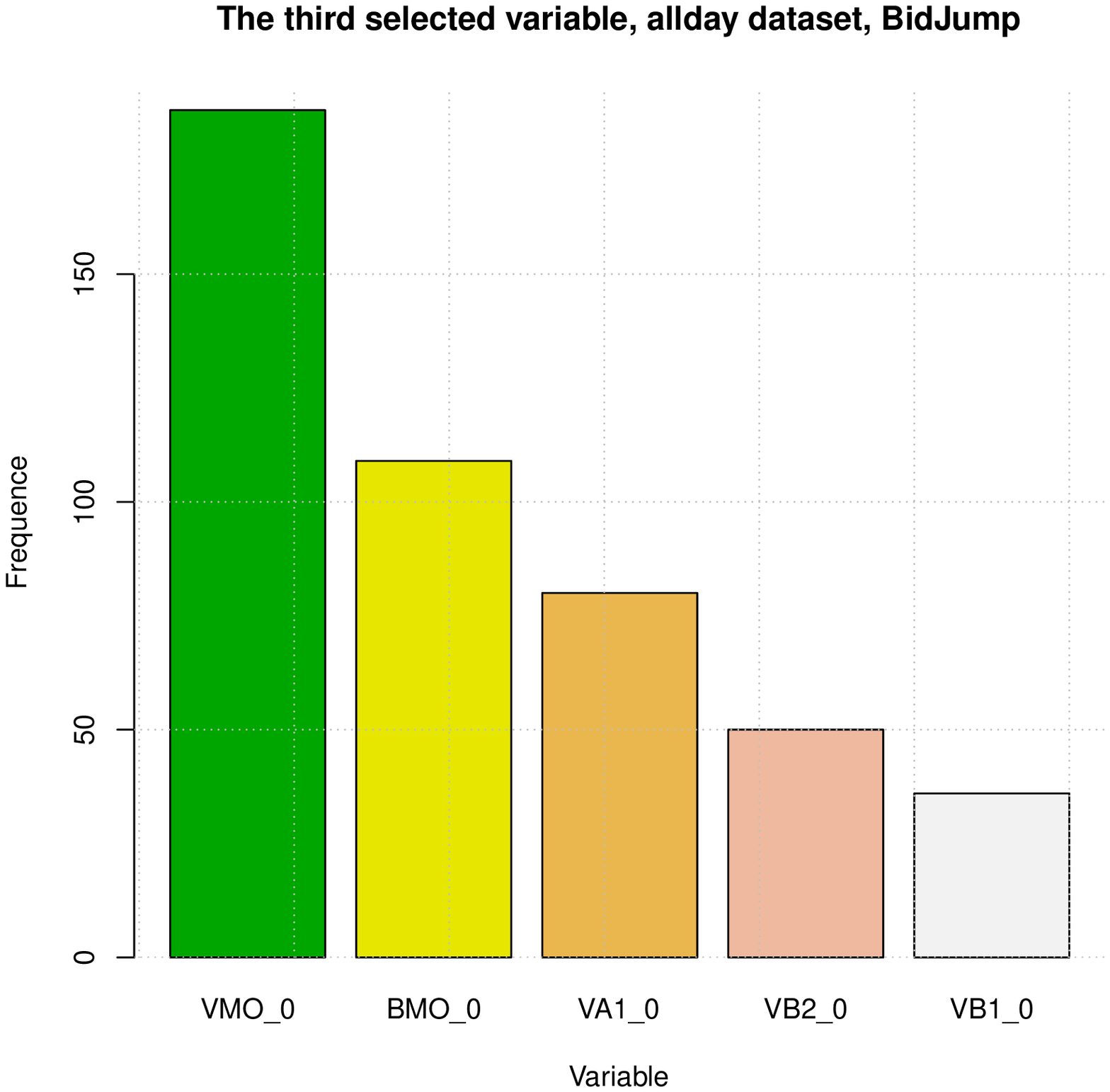}
\includegraphics[trim=0mm 0mm 0mm 0mm, clip, height=6cm, width=6cm, angle=-90]
{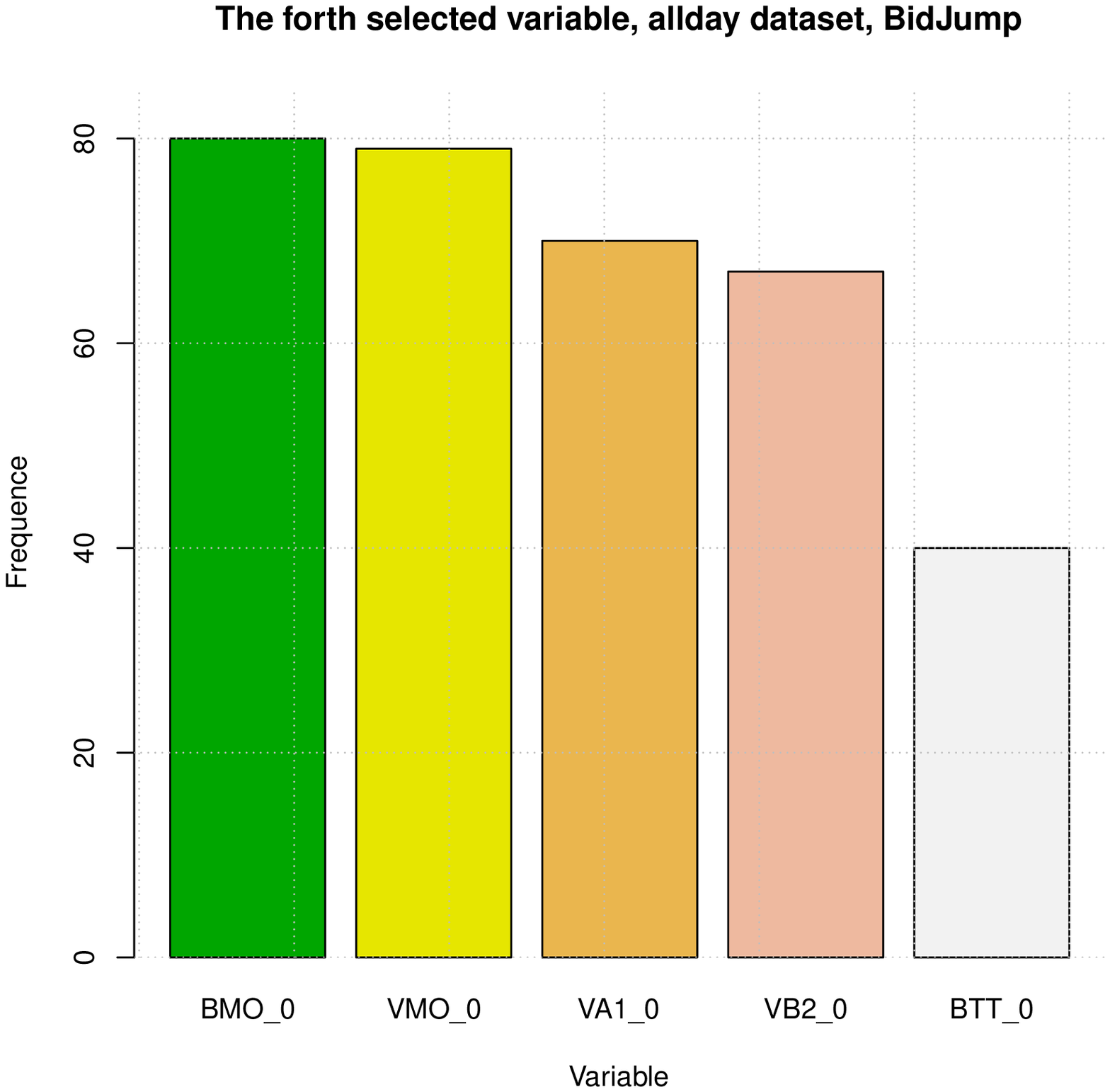}\\
\includegraphics[trim=0mm 0mm 0mm 0mm, clip, height=6cm, width=6cm, angle=-90]
{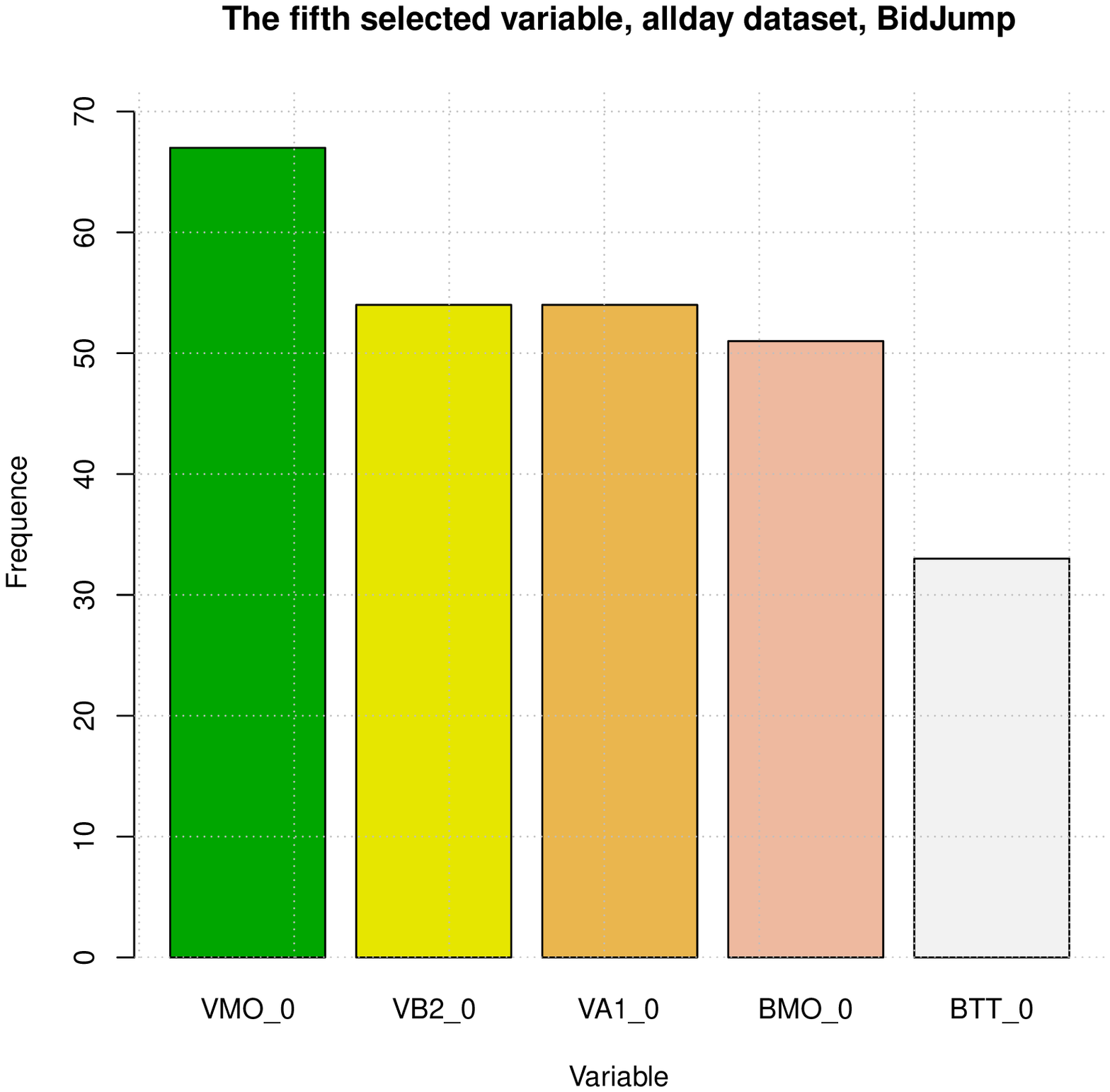}
\label{SelectedVariable_BidJump}
\end{figure}

\begin{figure}[!h]
\caption{Variable selection for AskJump prediction, CAC40, April, 2011. From left to right, from top to bottom, each figure shows how many times a variable is selected as the $k^{th}$ selected variable by $LASSO$, $k=\{1, \dots, 5\}$.}
\center
\includegraphics[trim=0mm 0mm 0mm 0mm, clip, height=6cm, width=6cm, angle=-90]
{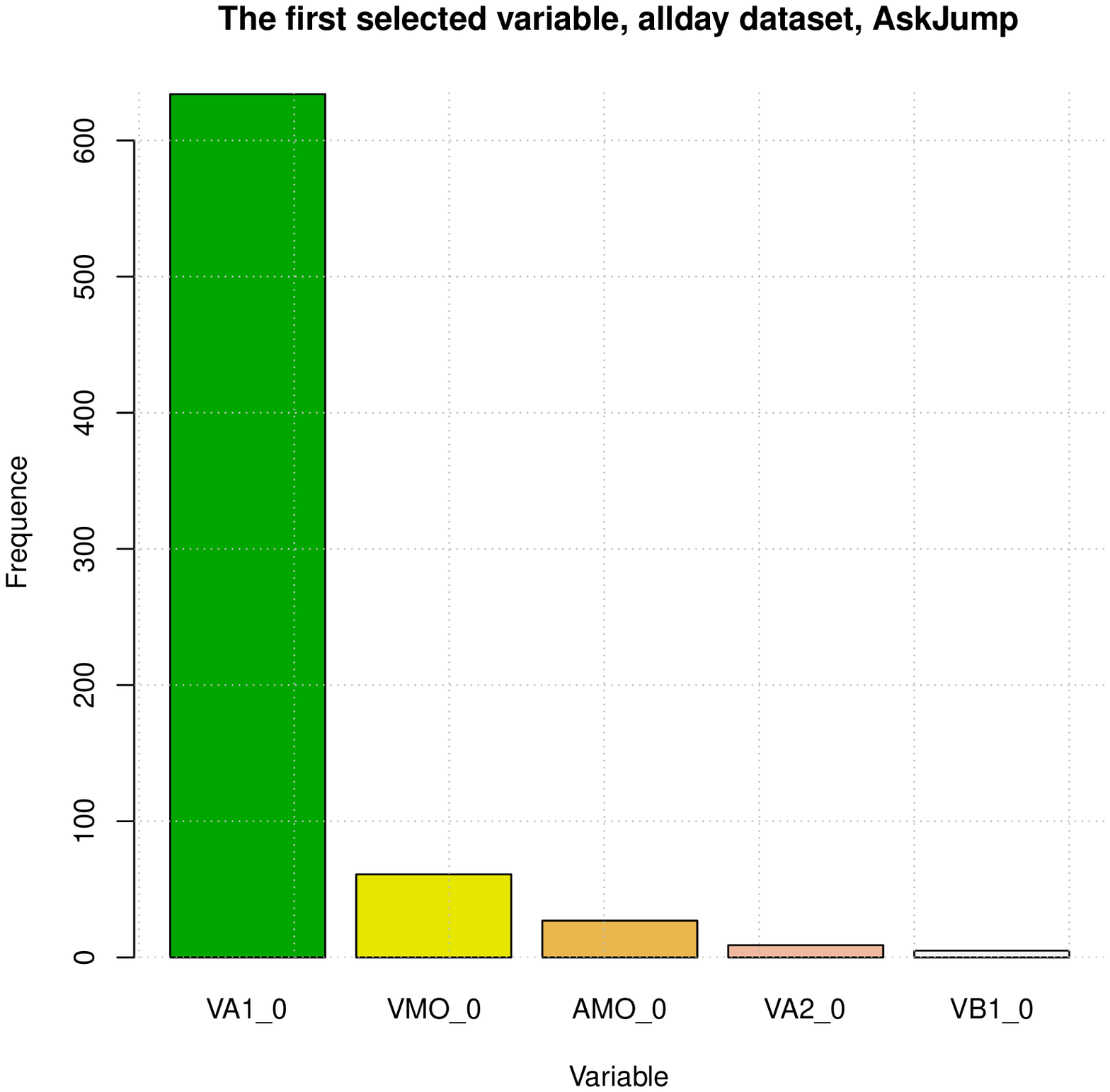}
\includegraphics[trim=0mm 0mm 0mm 0mm, clip, height=6cm, width=6cm, angle=-90]
{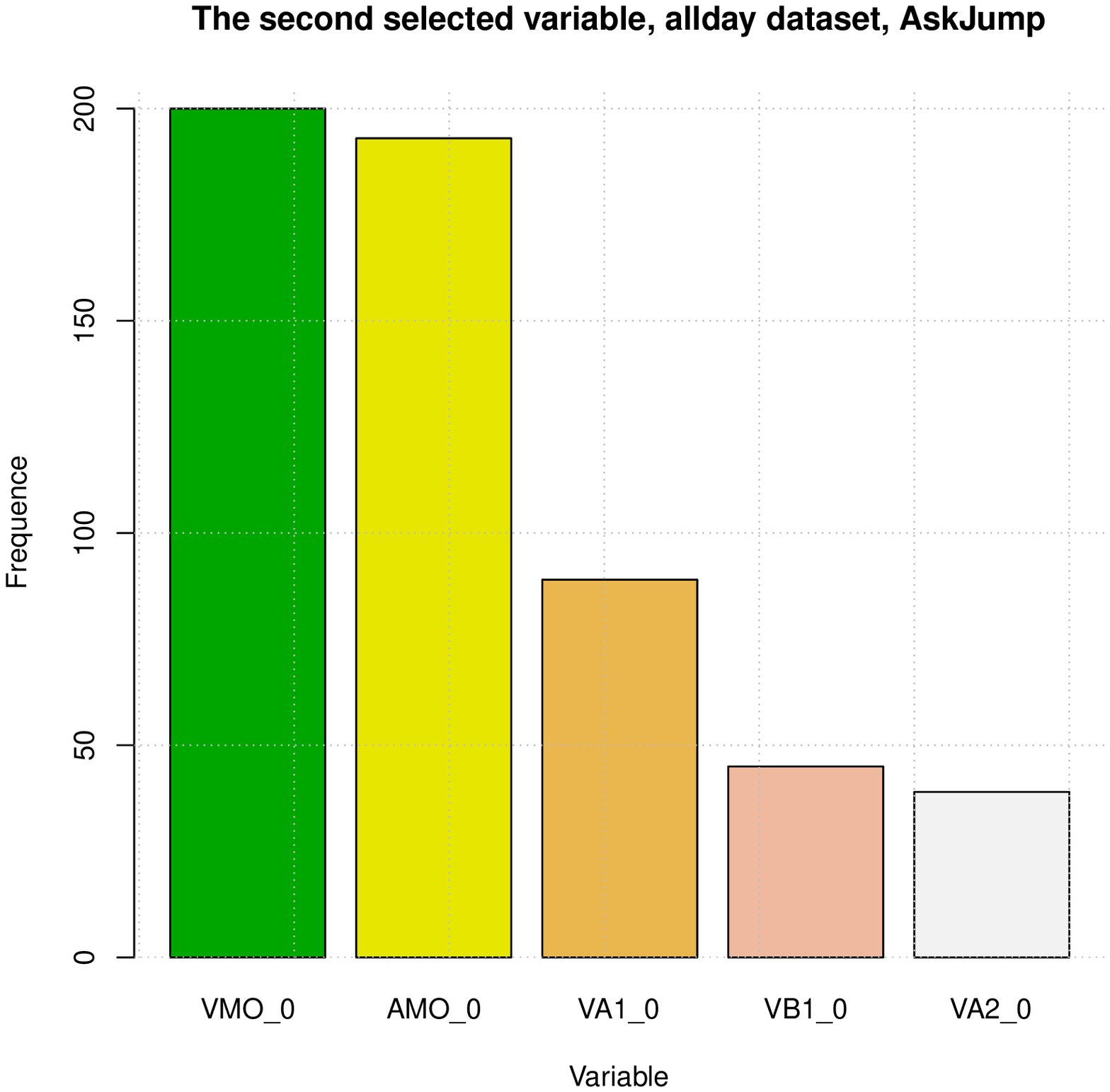}\\
\includegraphics[trim=0mm 0mm 0mm 0mm, clip, height=6cm, width=6cm, angle=-90]
{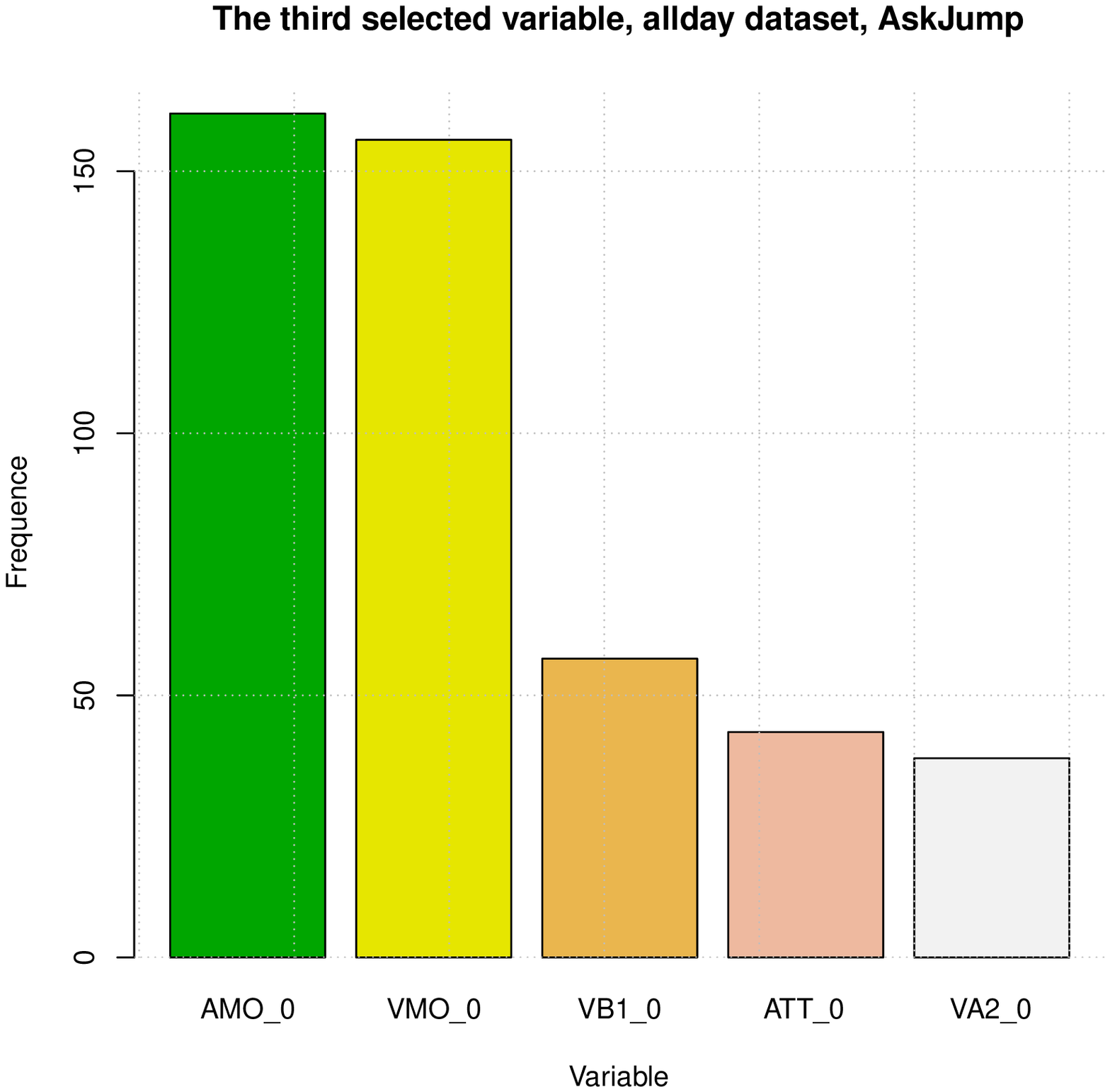}
\includegraphics[trim=0mm 0mm 0mm 0mm, clip, height=6cm, width=6cm, angle=-90]
{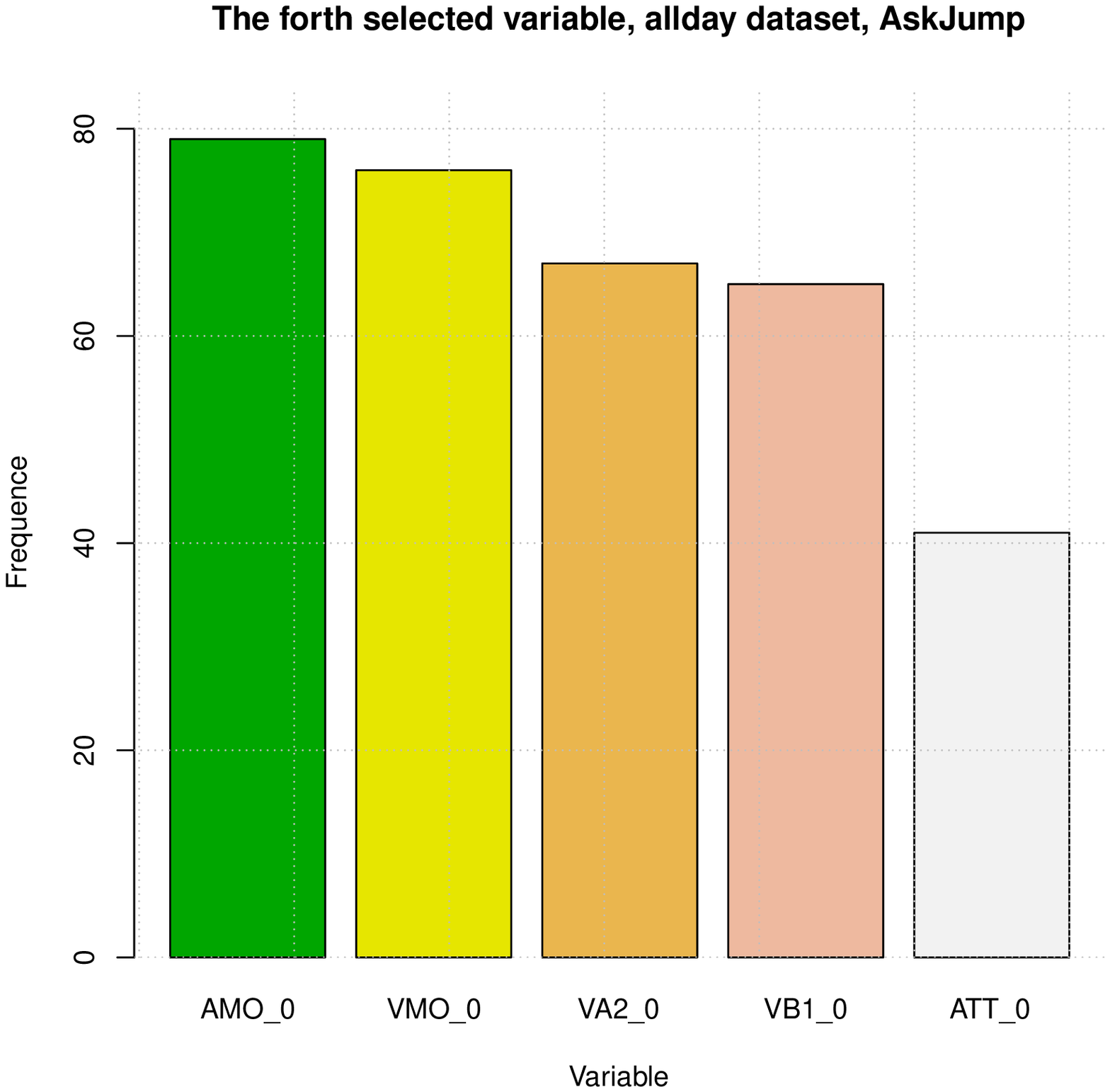}\\
\includegraphics[trim=0mm 0mm 0mm 0mm, clip, height=6cm, width=6cm, angle=-90]
{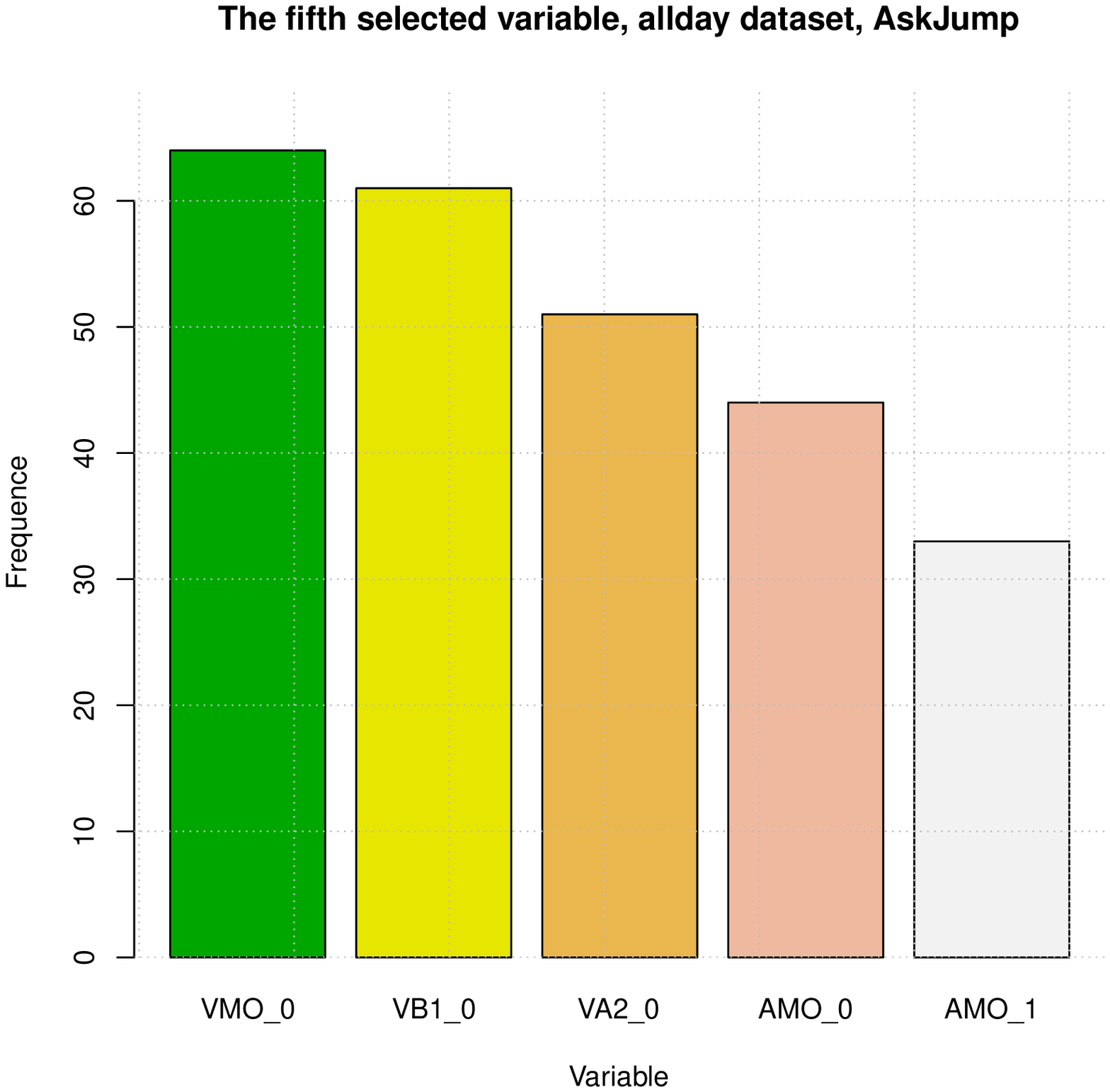}
\label{SelectedVariable_AskJump}
\end{figure}

Figure~\ref{SelectedVariable_BidJump} and~\ref{SelectedVariable_AskJump} show how many times a variable is selected as the first (second, third, forth, fifth) selected variable by $LASSO$. We denote the $i$ events lagged log volume on the $j^{th}$ bid (ask) limit price by $VBi\textunderscore{j}$ ($VAj\textunderscore{i}$). Similarly, $i$ events lagged log market order volume is denoted by $VMO\textunderscore i$ and $i$ events lagged binary variables are denoted by $BMO\textunderscore i$, $AMO\textunderscore i$, $BTT\textunderscore i$, $ATT\textunderscore i$ etc.  For the sake of simplicity, for each selection order $i$ ($i \in \{1, \dots, 5\}$), we show the frequency distribution of the five most frequently selected variables among $746$ backtests in each figure.

We observe that $VB1_0$, $BMO_0$ and $VMO_0$ are the most selected variables for predicting the future bidside \emph{inter-trade price jump} and that $VA1_0$, $AMO_0$ and $VMO_0$ are the most selected variables for predicting the future askside \emph{inter-trade price jump}. In contrast, \emph{trade-through} is less informative and contributes few to the price jump prediction. It implies that the market order is sensitive to the liquidity on the best limit price. As soon as the liquidity on the best limit price becomes significantly low, the next market order may touch it immediately. The information provided by $BMO_0$ ($AMO_0$) and $VMO_0$ recalls the phenomena of long memory of order flow, see Bouchaud et al. (2008) {~\cite{BFL2008}}. When a trader tries to buy or sell a large quantity of assets, he may split it into small pieces and execute them by market order successively. Consequently, precedent market order direction contributes to predict the next market order event.

\section*{Conclusion}

In this paper, we provide an empirical result on the relationship between bid-ask limit order liquidity balance and trade sign and an analysis on the prediction of the \emph{inter-trade price jump} occurrence by logistic regression. We show that limit order liquidity balance on best bid/best ask is informative to predict the next market order's direction. We then use limit order volumes, limit order price gaps and market order size to construct limit order book's feature for the prediction of \emph{inter-trade price jump} occurrence. LASSO logistic regression is introduced to help us identify the most informative limit order book features for the prediction. Numerical analysis is done on two separated datasets : morning dataset and afternoon dataset. LASSO logistic regression gets very good prediction results in terms of AUC value. The AUC value is consistently high on both datasets and all stocks whatever the liquidity is. This good prediction quality implies that limit order book profile is quite informative for predicting the incoming market order event. The variable selection by LASSO logistic regression shows that several variables are quite informative for \emph{inter-trade price jump} prediction. The trade sign and market order size and the liquidity on the best limit prices are the most informative variables. Nevertheless, the aggressiveness of market order, measured by \emph{trade-through}, has less important impact than we had expected. These results confirm that the limit order book is quite sensitive to the liquidity on the best limit prices and there is a long memory of order flow like what is shown by other authors. This paper is merely a first attempt to discover the information hidden in limit order book and further studies will be needed to understand better the full dynamics of limit order book.

\newpage
\section*{Appendix}
\begin{landscape}
\begin{figure}
\caption{The conditional probability of TradeSign vs bid-ask volume ratio, $Depth=1$, CAC40 stocks, April, 2011.}
\center
\includegraphics[trim=0mm 0mm 0mm 0mm, clip, height=8.0cm, width=8.0cm, angle=0]{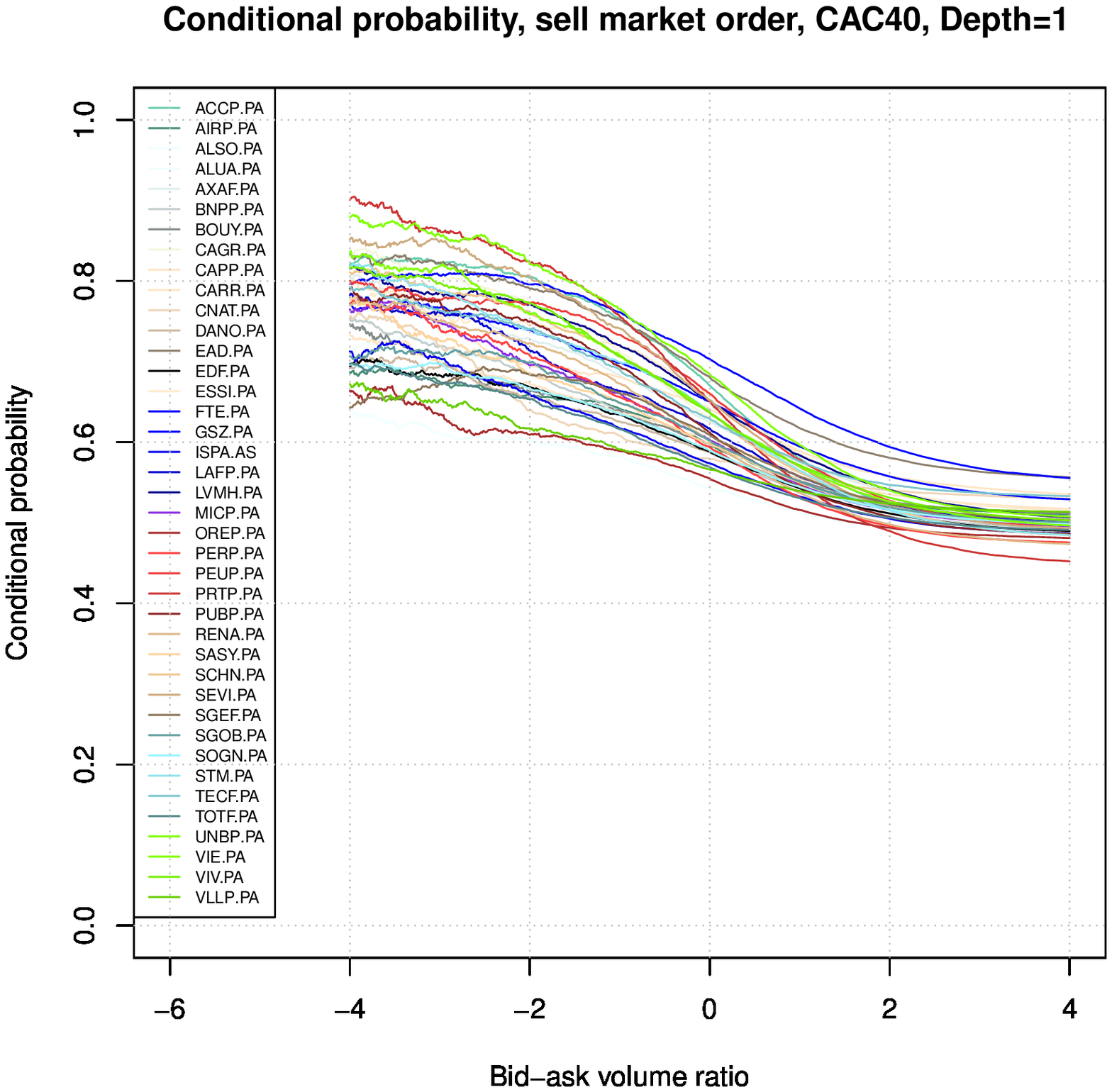} 
\includegraphics[trim=0mm 0mm 0mm 0mm, clip, height=8.0cm, width=8.0cm, angle=0]{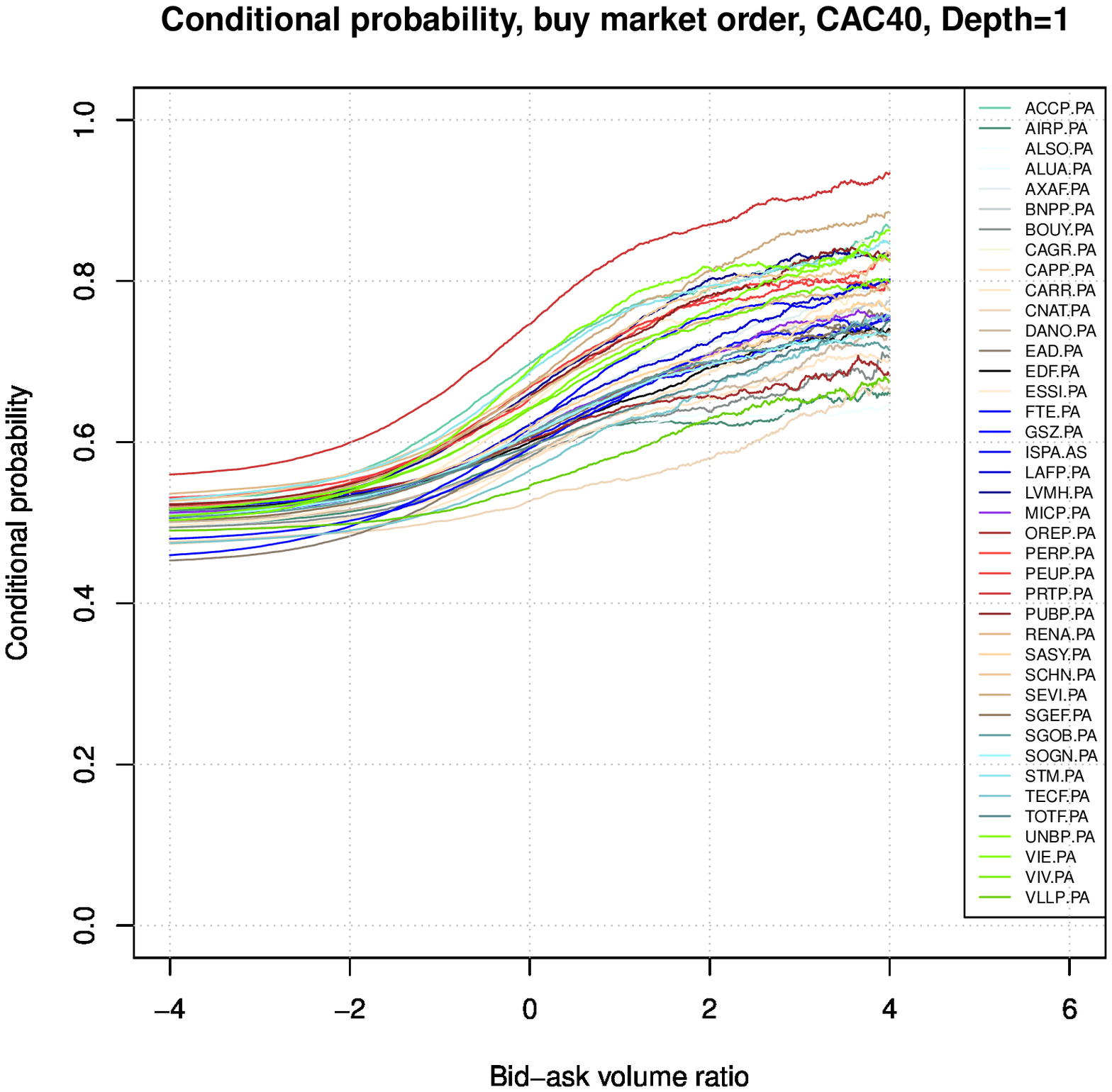} \\
\label{Figure_Vlm_BMO_CAC40}
\end{figure}
\end{landscape}
\newpage
\bibliographystyle{plain}
%\nocite{*}
\bibliography{Bib}

\end{document}